\documentclass[%
 reprint,
superscriptaddress,
 amsmath,amssymb,
 aps,prb,
]{revtex4-1}

\usepackage{graphicx}
\usepackage{dcolumn}
\usepackage{bm, color}
\usepackage{multirow}
\usepackage[T1]{fontenc}

\begin{document}%

\preprint{APS/123-QED}

\title{Phase diagrams for quantum Brownian motion in two-dimensional Bravais lattices}

 \author{Grace H. Zhang}
\affiliation{Department of Physics, Massachusetts Institute of Technology, Cambridge, Massachusetts 02139, U.S.A.}%
\affiliation{Department of Physics, Harvard University, Cambridge, Massachusetts 02138, U.S.A.}

\date{\today}

\begin{abstract}
We study quantum Brownian motion (QBM) models for a particle in a dissipative environment coupled to a periodic potential. We review QBM for a particle in a one-dimensional periodic potential and extend the study to that for a particle in two-dimensional (2D) periodic potentials of four Bravais lattice types: square, rectangular, triangular (hexagonal), and centered rectangular. We perform perturbative renormalization group analyses to derive the zero temperature flow diagrams and phase boundaries for a particle in these potentials, and observe localization behavior dependent on the anisotropy of the lattice parameters.  
\end{abstract}

\pacs{74.25.Dw}

\maketitle

\section{Introduction} \label{intro}

Quantum Brownian motion (QBM) models of a single particle in a dissipative environment were first studied to elucidate the phenomenon of quantum tunneling in macroscopic systems, where coupling to the environment is non-negligible and strong enough to damp classical motion \cite{Caldeira}. Followup works have focused on various classes of dissipative systems. These include two-level systems \cite{Chakravarty, Eckern}, which are becoming increasingly relevant in the context of quantum computation \cite{Makhlin}, and systems coupled to periodic potentials \cite{Fisher}, which have been used to study resonant tunneling in various condensed matter systems \cite{Kane, Kane2, Kane3, Furusaki, YiShort, YiLong}. Studies of both system types have revealed that increasing dissipation strength decreases the tunneling rate between barriers and induces a transition from a diffusive phase to a localized phase\cite{Schmid, Guinea, Chakravarty, Fisher}. 

In this paper, we investigate this transition between the diffusive and localized phases for systems in periodic potentials. We begin in Sec.~\ref{motivation} by briefly illustrating the physical and mathematical motivation for studying the QBM model. Then in Sec.~\ref{1D_Review}, we closely follow Refs.~\onlinecite{Schmid, Fisher} to review, respectively, the duality mapping between the weak and strong coupling regimes of 1D periodic systems and the corresponding renormalization group (RG) analysis, which together elucidate the diffusive-to-localized phase transition under increasing dissipation strength. The new contribution of this paper starts in Sec.~\ref{2D_Uncoupled} and examines a particle in a dissipative environment coupled to two-dimensional (2D) periodic potentials. Previous studies have focused on isotropic lattice potentials with permutation symmetry among all coordinate axes \cite{YiShort, YiLong}. Here, we generalize the study to periodic potentials with space groups of among four types of 2D Bravais lattices: square, rectangular, hexagonal (triangular), and centered rectangular. We apply RG analysis in the perturbative regime and the strong coupling regime for the real space lattice and its dual reciprocal space lattice to obtain the corresponding flow diagrams. 

We find that depending on the system lattice parameters, the particle can exhibit the diffusive delocalization phase, the point-localization phase, or a line-localization phase, and obtain the corresponding phase boundaries as a function of the lattice anisotropy. These results may also have implications in other systems onto which the 2D QBM model can be mapped. Specifically, multi-dimensional QBM models have been used to study resonant tunneling in multi-lead quantum dot systems under the effect of Coulomb blockade and the multi-channel Kondo problem in the multi-dimensional generalization of the Toulouse-limit \cite{YiShort, YiLong}, the charge and spin transport of Luttinger liquids through different types of barrier structures \cite{Kane, Kane2, Kane3}, as well as related problems in condensed matter and statistical physics \cite{Affleck, Beri, Altland, Teo}.

\section{Quantum dissipation through a linear coupling term} \label{motivation}

In this section, we very briefly illustrate the physical origin of a linearly coupled dissipative term following Refs.~\onlinecite{simons_lec, QCMFT}. Additionally, we note that $\hbar =1$ in all calculations to follow. 

\subsection{Physical motivation: averaging over fluctuations in the environment} \label{partString}

\begin{figure}[b]
\includegraphics[width=0.8\columnwidth]{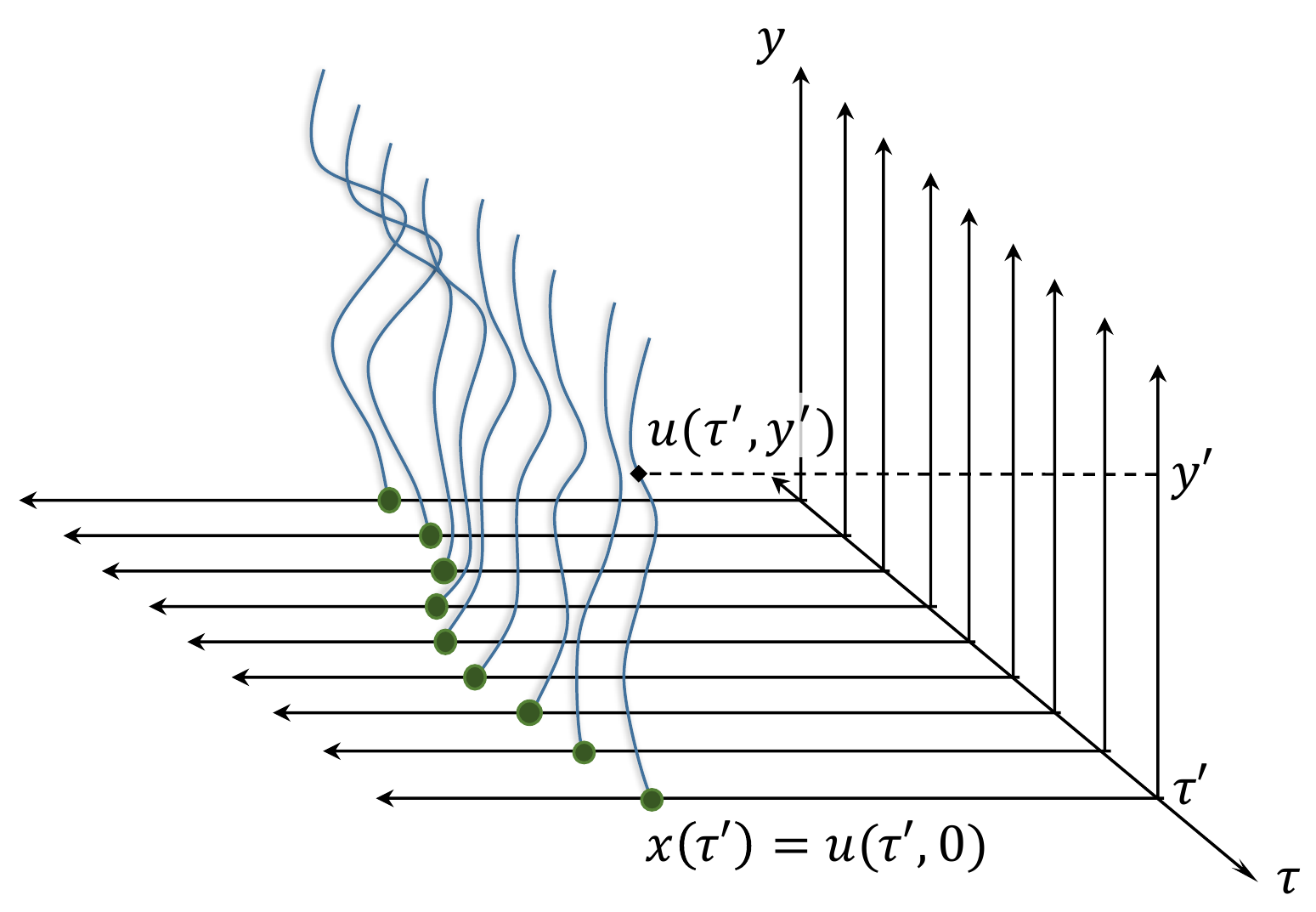}
\caption{\label{string} Schematic of a particle coupled to one end of a fluctuating string. The effects of the motion of the string on the particle are averaged over to obtain an effective Euclidean action $S_{\text{eff}}$ whose only degrees of freedom are those of the particle. }\end{figure}

Let us consider a system that consists of a single particle, whose position on the $x$-axis $x(\tau)$ varies as a function of imaginary time $\tau$, and a string, which extends along the $y$-direction and whose transverse fluctuations in time are indicated by $u(\tau, y)$ (see Fig.~\ref{string}). The particle is coupled to the end of the fluctuating string at $y=0$. We are mainly interested in the behavior of the particle and thus trace over all fluctuations of the string to obtain an effective action, where effects of the string have been averaged out and the only degrees of freedom remaining are those of the particle. The string in this model represents an external environment to which the object of interest couples. 

The action of the string and the particle are, separately,
\begin{equation}
\begin{aligned}
S_{s} 	&= \int d\tau \int dy \Bigg[ \frac{\rho}{2}\dot u^2 + \frac{\sigma}{2}\Big(\frac{\partial u}{\partial y} \Big)^2 \Bigg] \\
S_{p}	&= \int d\tau \Bigg[ \frac{m}{2} \dot x^2 + U[x(\tau)] \Bigg], \\
\end{aligned}
\end{equation}
where $\rho$ and $\sigma$ denote the density and tension of the string, respectively, and $U[x]$ indicates some potential that the particle experiences depending on its position $x$. The coupling condition can be imposed on the overall partition function $Z$ through the $\delta$-function,
\begin{equation}
Z = \int Du(\tau, y) \int Dx(\tau) \delta\Big( x(\tau) - u(\tau, 0)\Big) e^{-S_{s}-S_{p}}.
\end{equation}
Writing the $\delta$-function as follows,
\begin{equation}
\delta(x(\tau) - u(\tau, 0)) = \int D f(\tau) e^{i \int d\tau f(\tau)(x(\tau) - u(\tau,0))},
\end{equation}
transforming to Fourier space, and performing the Gaussian integral over $\tilde u(\omega, k)$ and subsequently over $\tilde f(\omega)$, we obtain an effective action $S_{\text{eff}}$ for the particle such that, 
\begin{equation}\label{Hp}
\begin{aligned}
Z 		&= \int D \tilde x(\omega) e^{-S_{ \text{eff} }}, \\
S_{\text {eff}} 	&= \frac{\eta}{2} \int \frac{d\omega}{2 \pi} |\omega| |\tilde x(\omega)|^2 + S_p,
\end{aligned}
\end{equation} 
where $\eta = 2 \sqrt{\rho \sigma}$. Transforming back to time space then gives us
\begin{equation}\label{Hp2}
S_{\text {eff}} = S_p + \frac{\eta}{2} \int d\tau d\tau' \Bigg(\frac{x(\tau)-x(\tau')}{\tau-\tau'}\Bigg)^2.
\end{equation}
Thus, we see that the string effectively induces a nonlocal friction term ($\sim \dot xx$) into the action of the particle, with $\eta$ as the real and positive friction coefficient. The string in this case can be generalized to an external environment with which the particle interacts. In general, we do not know the precise details of a system's dissipation mechanism, and the value of the phenomenological friction coefficient $\eta$ is extracted from experimental results. 

\subsection{Mathematical motivation: available RG technology from the 2D sine-Gordon model}

The quantum dissipation model in 1D is an analogue of the 2D sine-Gordon model. Although this point will be covered in more detail in later sections, we briefly point out the correspondence here. From Eq.~(\ref{Hp}), assuming that $S_p$ contains the standard kinetic energy term $\sim \omega^2|\tilde x(\omega)|^2$, in the long-wavelength limit, the dominant contribution to $S_{\text {eff}}$ comes from the dissipative term $\sim |\omega||\tilde x(\omega)|^2$. Thus, the two-point correlation function $\langle \tilde x(\omega) \tilde x(-\omega) \rangle \sim |\omega|^{-1}$. As a result, integrals of the form $\int d \omega \langle \tilde x \tilde x \rangle$ in perturbative analyses lead to the emergence of logarithmic interactions at large distances. This is in correspondence with the 2D sine-Gordon model,
\begin{equation} \label{HSG}
\beta H_{\text {SG}} = -\int d^2 x \bigg[ \frac{J}{2} (\nabla \theta(x))^2 + h_p \cos(p\theta(x)) \bigg],
\end{equation}
where $J$ is the inverse-temperature coupling coefficient and $h_p \cos(p\theta(x))$ represents a symmetry breaking field. 
In the long-wavelength limit, Eq.~(\ref{HSG}) is dominated by ${\sim q^2|\tilde \theta(q)|^2}$, which, through $\int d^2q \langle \tilde \theta \tilde \theta \rangle$ in perturbative analyses, also generates a logarithmic interaction potential \cite{Kardar}. Thus, the respective first terms of the effective action in Eq.~(\ref{Hp}) and the 2D sine-Gordon Hamiltonian in Eq.~(\ref{HSG}) both lead to logarithmic interactions under perturbative analyses.

As such, the perturbative RG techniques used for analyzing the 2D sine-Gordon model, which describes the roughening transition, can be applied to the 1D QBM model. The resulting scaling relations of the two problems can then be mapped onto each other (as shown later in Sec. \ref{1DRG}). These connections are crucial for the calculations carried out in the rest of this paper. 

\section{Review of QBM in a 1D periodic potential} \label{1D_Review}

For a particle in a periodic potential of periodicity $x_0$, we rewrite Eq.~(\ref{Hp2}) as 
\begin{equation} \label{Sinu}
\begin{aligned}
S_0 &= \frac{\eta}{4 \pi } \int d\tau \int d\tau' \Bigg[ \frac{x(\tau) - x(\tau')}{\tau - \tau'} \Bigg]^2 + \int d\tau \frac{m}{2}\dot x ^2 (\tau) \\
S_1 &= -V \int d \tau \cos \Bigg[ \frac{2 \pi x(\tau)}{x_0} \Bigg].
\end{aligned}
\end{equation}
We make an important note here before proceeding. In this paper, as well as in previous works on QBM in periodic potentials \cite{YiLong, YiShort, Fisher, Schmid, Guinea}, we consider sinusoidal potentials such as that in $S_1$ of Eq.~(\ref{Sinu}) and analyze their relevance conditions to derive the system phase boundaries. Consideration of a generic periodic potential then involves decomposing it into its Fourier series and retaining the most relevant sinusoidal terms. 

To proceed with calculations, we first adapt the notation in Ref.~\onlinecite{Fisher} and scale our parameters to be dimensionless as follows, 
\begin{equation}
\begin{aligned}
\alpha &= \frac{\eta x_0^2}{2\pi } \\
V_0 &= \frac{V}{\Lambda} \\
\phi(\tau) &= \frac{2 \pi x(\tau)}{x_0},
\end{aligned}
\end{equation}
where $\Lambda$ is the high frequency cutoff corresponding, in this case, to the quantum mechanical energy $\frac{(2 \pi)^2}{m x_0^2}$ required to confine a particle of mass $m$ within a lattice spacing of $x_0$.  
Transforming to Fourier space, we arrive at the scaled action,
\begin{equation} \label{scaledH}
\begin{aligned}
S_0 &= \frac{1}{2} \int \frac{d\omega}{2 \pi} \tilde K(\omega) |\tilde \phi(\omega)|^2, \quad \tilde K(\omega) = \frac{\alpha}{2 \pi} |\omega| + \frac{\omega^2}{\Lambda} \\
S_1 &= - V_0 \Lambda \int d\tau \cos \big( \phi(\tau) \big).
\end{aligned}
\end{equation}

\subsection{Duality Mapping}

In this section, we follow Ref.~\onlinecite{Schmid} to review the duality mapping between the weak and strong potential regimes of the system. The self-duality of 1D QBM in a periodic potential then identifies the existence of a diffusive delocalized phase and a localized phase, as well as the corresponding critical point of transition. 
\subsubsection{Weak coupling expansion}

In the weak coupling regime, expansion of the partition function $Z = \int D \phi(\tau) e^{-S_0} e^{-S_1}$ leads to a power series summing over multi-particle charge configurations describing neutral plasmas. To see this, we proceed by assuming $V_0 \ll 1$ and expanding $e^{-S_1} $ to observe its weight contribution to the partition function $Z$,
\begin{equation}
\begin{aligned}
		&\exp \Big( V_0 \Lambda \int d\tau \cos (\phi(\tau)) \Big) \\
		=&\sum_n \frac{(V_0 \Lambda/2)^n}{n!} \int d\tau_1 \ldots d\tau_n \\
		& \qquad  \times \sum_{\{ e_j = \pm 1\} } \exp\Big[i \int d\tau \sum_{j=1}^n e_j \delta(\tau-\tau_j) \phi(\tau)\Big].
\end{aligned}
\end{equation}
We note here that in performing the path integral, we can separate out the zero-frequency mode, that is, 
\small
\begin{equation}
\begin{aligned} 
\int & D\phi(\tau) e^{-S} = \sum_n \frac{(V_0 \Lambda/2)^n}{n!} \int \prod_{j=1}^{n}  d\tau_j \sum_{\{ e_j = \pm 1\} } A \Big [\tau, \{e_j, \tau_j\} \Big ],
\end{aligned}
\end{equation}
\normalsize
where $\{e_j, \tau_j\} $ indicates a particular plasma configuration consisting of charges $\{e_j\}$ located at times $\{ \tau_j \}$, and 
\begin{equation} \label{zero_cont}
\begin{aligned}
A \Big [\tau, & \{e_j, \tau_j\} \Big ] \\
= & \int \prod_{\omega=0}^\infty d \tilde \phi(\omega) \exp\Bigg\{-\sum_\omega \Bigg[ \frac{\tilde K(\omega)}{2} |\tilde \phi(\omega)|^2 \\
& \qquad -i \sum_{j=1}^n e_j \int d\tau \delta(\tau-\tau_j) e^{-i\omega\tau} \tilde \phi(\omega)\Bigg] \Bigg\}\\
	=& \int d \tilde \phi_0 \exp \Big[ i \sum_{j=1}^n e_j \int d\tau \delta(\tau-\tau_j) \tilde \phi_0 \Big] \prod_{\omega \neq 0} d \tilde \phi(\omega) \ldots,
\end{aligned}
\end{equation}
where we have denoted $\tilde \phi_0 = \tilde \phi(\omega=0)$ and used the fact that $\tilde K(0)=0$, as seen from Eq.~(\ref{scaledH}) ${(\tilde K(\omega) = \frac{\alpha}{2 \pi} |\omega| + \frac{\omega^2}{\Lambda} )}$. Looking at the integral over the zero-frequency mode in the last line of Eq.~(\ref{zero_cont}), we see that configurations where $\sum_{j=1}^n e_j \neq 0$ do not contribute to the partition function. In other words, identifying each $e_j$ as the charge of a classical particle and $\sum_{j=1}^n e_j$ as the total charge of a plasma of $n$ such particles, only neutral plasma configurations are allowed. Thus, $n=2N$, and we can identify a charge density 
\begin{equation}
\rho_{2N}(\tau) = \delta(\tau - \tau_1) - \delta(\tau - \tau_2) + \delta(\tau - \tau_3) - \ldots -\delta(\tau - \tau_{2N}) 
\end{equation}
that corresponds to a neutral plasma of $n$ particles with $(2N)!/N!N!$ distinct configurations originating from the different ways of distributing $N$ $(+)$ particles and $N$ $(-)$ particles on $2N$ $\{ \tau_j \}$ sites. We then obtain the expression 
\begin{equation}
\begin{aligned}
e^{- S_1} =&\sum_N \Big(\frac{(V_0 \Lambda/2)^N}{N!}\Big)^2 \int d\tau_1 \ldots d\tau_{2N} \\
		& \qquad  \qquad \times \sum_{\{ e_j = \pm 1\} } \exp\Big[i \int d\tau \rho_{2N}(\tau) \phi(\tau)\Big],
\end{aligned}
\end{equation}
which identifies $\rho(\tau)$ as a source term and puts the integrand of the partition function $Z$ into Gaussian form in $\phi(\tau)$. Integrating over the fluctuations in $\phi(\tau)$, we arrive at the following,
\begin{equation} \label{Zweak}
\begin{aligned}
Z &= \sqrt{\bar D} \sum_N \Big(\frac{(V_0 \Lambda/2)^N}{N!}\Big)^2 \int d\tau_1 \ldots d\tau_{2N} \\
&\times \exp\Bigg[ -\frac{1}{2} \int d\tau d\tau' \rho_{2N}(\tau) K^{-1}(\tau-\tau')\rho_{2N}(\tau') \Bigg],
\end{aligned}
\end{equation}
where $\bar D =  \prod_{\omega \neq 0} \tilde K(\omega)$ and, as calculated in Ref.~\onlinecite{Schmid}, 
\begin{equation} \label{weak}
\begin{aligned}
  \bar D^{-1}(\tau) &= \int \frac{d\omega}{2\pi} \tilde K^{-1}(\omega) e^{-i\omega(\tau)} \\
  		&=\begin{cases}
               -\frac{\Lambda}{2}|\tau|,  	&\alpha \Lambda |\tau| \ll 1\\
		-\frac{2}{\alpha}\ln \Big(\frac{\alpha \Lambda |\tau|}{2 \pi} \Big), 	 &\alpha \Lambda |\tau| \gg 1,
            \end{cases}
\end{aligned}
\end{equation}
where we note the logarithmic behavior of the interaction at long distances.

\subsubsection{Strong coupling expansion} \label{instanton}

In the limit of $V_0 \gg 1$, the tunneling of the quantum particle between adjacent minima of the periodic potential can be well described, using the WKB approximation, by its classical undamped path in the inverted potential. Denoting $S' = S - \int \frac{d\omega}{2 \pi} \frac{\alpha}{4 \pi} |\omega| |\phi(\omega)|^2$ as the full undamped action, we solve the following corresponding differential equation, 
\begin{equation} \label{instantonDiff}
\ddot {\bar {\phi}}(\tau) - \Lambda^2 V_0 \sin(\bar \phi(\tau))=0.
\end{equation}
We arrive at the single instanton/anti-instanton solution described by 
\begin{equation} \label{single_inst}
\bar \phi(\tau) = \pm f(\tau) = \pm 4 \arctan(e^{\omega_0 \tau}), 
\end{equation}
where $\omega_0 = \sqrt{\Lambda^2 V_0}$. Substituting $f(\tau)$ back into $S'$, we calculate the energy of a single instanton/anti-instanton $s = 8 \sqrt{V_0}$ which is $\gg 1$ in the strong coupling regime. 
A general solution is then a combination of instantons and anti-instantons distributed at different points in time,
\begin{equation} \label{inst_gas}
\bar \phi_n (\tau) = \sum_{j = 1}^n e_j f(\tau - \tau_j),
\end{equation}
where $e_j = 1$ or $e_j = -1$ describes an instanton or anti-instanton, as long as the time separation between any two instantons is substantially larger than the width of a single instanton ($|\tau_j-\tau_k| \gg w_0^{-1}$ for $j \neq k$).  Next, denoting $h(\omega)$ as the Fourier transform of $\dot f(\tau)$, we also note the following property of $\tilde \phi(\omega)$ resulting from the convolution theorem for Fourier transforms, 
\begin{equation}
-i \omega \tilde \phi(\omega) = h(\omega) \sum_j e_j e^{i \omega \tau_j} \label{hphi}.
\end{equation}
Then, substituting $\phi_n$ into the full action $S$, we obtain the following expression
\begin{equation}
S[\phi_n] = ns + \frac{1}{2} \sum_{jk} e_j e_k \Delta(\tau_j - \tau_k), 
\end{equation}
where the second term describes an effective pairwise interaction between the instantons and anti-instantons, and
\begin{equation} \label{strong}
\begin{aligned}
  \Delta(\tau) &= \frac{\alpha}{2 \pi} \int \frac{d\omega}{2\pi} \frac{| h(\omega)|^2}{|\omega|} e^{-i\omega(\tau)} \\
  	&= \begin{cases}
               -c \frac{\alpha}{2 \pi} \omega_0^2 \tau^2,  	&\omega_0 |\tau| \ll 1\\
		-2 \alpha \ln (\omega_0 \tau), 	 		&\omega_0 |\tau| \gg 1,
            \end{cases}
\end{aligned}
\end{equation}
where $c$ is a constant of order unity. First, we note that since $h(\omega = 0) = 2\pi$ is a constant, integration over the zero-frequency mode leads to a divergence in energy unless $\sum_j e_j = 0$. Therefore, only solutions with equal numbers of instantons and anti-instantons are allowed, imposing the condition $n=2N$ as in the case of the weak coupling expansion. 

Next, taking into account Gaussian fluctuations around $\bar \phi$, the work for which is not shown here but can be found in various reference texts\cite{QCMFT}, the fugacity $e^{-s}$ is replaced with $\omega_0 (2s/\pi)^{1/2} e^{-s}$. Thus, identifying $\rho_{2N}(\tau)$ again as the density of a ``neutral'' mixture of instantons and anti-instantons, we arrive at the following expression for the partition function, 
\begin{equation} \label{Zstrong}
\begin{aligned}
Z &= \sum_N \Bigg(\frac{(\omega_0 e^{-s} \sqrt{2 s / \pi})^N}{N!}\Bigg)^2  \int d\tau_1 \ldots d\tau_{2N} \\
&\times \exp\Bigg[ -\frac{1}{2} \int d\tau d\tau' \rho_{2N}(\tau) \Delta(\tau-\tau')\rho_{2N}(\tau') \Bigg]. 
\end{aligned}
\end{equation} 

\begin{figure}[h]
\includegraphics[width=0.8\columnwidth]{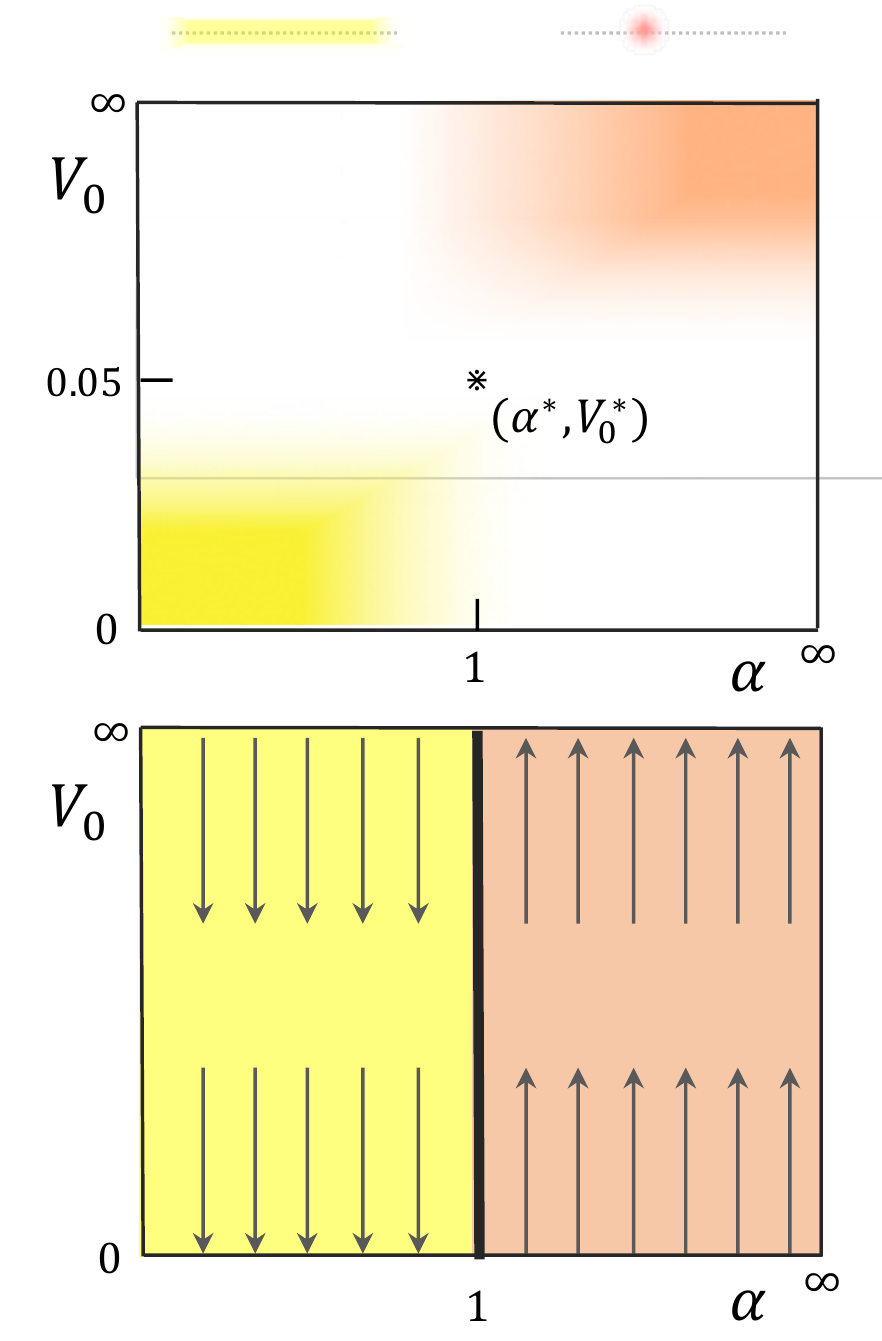}
\caption{\label{schm} Top: Duality mapping between the weak and strong potential coupling regimes of the 1D QBM model reveals a fixed point at $\alpha^* = 1$, $V_0^*=0.05$ \cite{Schmid}. As the parameters are increased past this point, the system transitions from a diffusive delocalized phase to a localized phase, represented by the schematics above the figure. Bottom: RG analysis clarifies that at zero temperature, this transition from the delocalized phase to the localized phase occurs at the same value of $\alpha^*$ for any nonzero initial height of the potential barriers $V_0$ \cite{Fisher}. } 
\end{figure}

Comparing the effective interaction potentials of the corresponding charge and instanton densities $\rho_{2N}(\tau)$ in Eqs.~(\ref{weak}) and (\ref{strong}), we see that just as in the weak coupling expansion, a logarithmic interaction is generated between the instantons and anti-instantons separated by large distances in time in the strong coupling expansion. If we neglect their behavior in the core regions (small $|\tau|$), we are able to map them onto each other through the relation,
\begin{equation} \label{a_map}
\begin{aligned}
\alpha & \rightarrow \frac{1}{\alpha}.
\end{aligned}
\end{equation}
Next, comparing the partition functions in Eqs.~(\ref{Zweak}) and (\ref{Zstrong}), we also recognize the mapping between the prefactors raised to the $n$th power in each of the power sums,
\begin{equation} \label{s_map}
\begin{aligned}
\frac{8}{\sqrt{\pi}} V_0^{\frac{1}{4}} e^{-s} &\rightarrow \frac{2 \pi}{\alpha} V_0 . 
\end{aligned}
\end{equation}
Mappings (\ref{a_map}) and (\ref{s_map}) then lead to the following fixed points, as shown in the top of Fig.~\ref{schm},
\begin{equation}
\begin{aligned}
\alpha^* &= 1 \\
V_0^* & \approx 0.05.
\end{aligned}
\end{equation}
However, while the weak coupling expansion did not assume conditions of the system parameters, the expansion from strong coupling assumed the use of the WKB approximation to be valid. Here, as $V_0^* \approx 0.05$, our fixed point is not in the strongly coupled regime where the latter condition is satisfied. Therefore,  the estimate of $\alpha^*$ is also not reliable enough, motivating further analysis using RG. 

\subsection{Perturbative RG analysis under a 1D periodic potential} \label{1DRG}

Before performing RG analysis, we point out that, crucially, the periodic potential imposes two constraints on our system. First, rescaling in the time coordinate does not affect the real space coordinate on which our particle rests. The periodic nature of our system is constrained by our system size in real space, which does not change under renormalization in time. Thus, $x_0$ stays the same under renormalization and is not treated as a scalable parameter but rather as a given system constant. Secondly, a sensible renormalization procedure requires that the dimensionless position function $\phi(\tau)$ maps back onto itself with the same periodicity $\phi \rightarrow \phi + 2\pi$, else breaking the symmetry of the system. To preserve this symmetry, the renormalization constant in $\phi'(\tau) = \phi(\tau)/z$ is $z=1$. 

An important consequence follows. Since $\phi(\tau)$ does not get renormalized, we can see from power counting that under rescaling $\omega \rightarrow \omega b^{-1}$, the quadratic term ($\sim \omega^2$) and any terms higher order in $\omega$ are irrelevant in the presence of the dissipation term in the 1D bare action. Thus, in the calculations and analysis that follow, we take the limit $\tilde K(\omega) \rightarrow \alpha |\omega| / 2\pi$. 

From here, we perform the standard renormalization procedure to study the effects of the perturbation terms in 
\begin{equation} \label{S_pert}
S = S_0 + \langle S_1 \rangle -\frac{1}{2}\Big( \langle S_1^2 \rangle - \langle S_1\rangle^2 \Big) + O(V_0^3).
\end{equation}
In Eq.~(\ref{S_pert}), $\langle \ldots \rangle$ indicates averaging over the fast modes of $\tilde \phi(\omega)$, which are identified through
\begin{equation}
 \tilde \phi(\omega)=\begin{cases} 
              \tilde \phi_<(\omega), \text{			for } |\omega| < \Lambda/b\\
		\tilde \phi_>(\omega), \text{		for } \Lambda/b \leq |\omega| < \Lambda,
            \end{cases}
\end{equation}
where $\Lambda$ is the high frequency cutoff first introduced in Eq.~(\ref{scaledH}). These modes are then identified in time space as
\begin{equation}
\begin{aligned}
\phi(\tau) 	&= \int_{0}^{\Lambda/b} \frac{d\omega}{2\pi} \tilde \phi(\omega) e^{i\omega\tau} + \int_{\Lambda/b}^{\Lambda} \frac{d\omega}{2\pi} \tilde \phi(\omega) e^{i\omega\tau} \\
		& = \phi_<(\tau) + \phi_>(\tau).
\end{aligned}
\end{equation}
The two-point correlation function of the fast modes is then
\begin{equation}
G(\tau) \equiv \langle \phi_>(\tau) \phi_>(0) \rangle = \int_{\Lambda/b}^{\Lambda} \frac{d\omega}{2\pi} \frac{e^{i\omega\tau}}{\tilde K(\omega)},
\end{equation}
which we will use to integrate out the corresponding fast fluctuations. 
In obtaining $G(\tau)$, a smoothing function,
\begin{equation}
W\left(\frac{\omega}{\Lambda/b}\right) = \frac{|\frac{\omega}{\Lambda/b}|}{\sqrt{1+\left( \frac{\omega}{\Lambda/b} \right)^2}},
\end{equation}
is incorporated into the integral above, where we note that $W\left(\frac{\omega}{\Lambda/b}\right) \approx 1$ when $\omega \gg \Lambda/b$ and vanishes slowly as $\omega \rightarrow \Lambda/b$, so as to prevent the generation of spurious long-range behavior in $G(\tau)$. Here, we will take the results as calculated previously in Ref.~\onlinecite{Fisher} for use in later calculations:
\begin{equation} \label{gtau}
\begin{aligned}
G(\tau) &= \frac{2}{\alpha}Y_0(\Lambda \tau /b) \quad \text{for } \Lambda\tau \gg 1, \\
G(0) &= \frac{2}{\alpha}\ln(b),
\end{aligned}
\end{equation}
where $Y_0(z)$ is the modified Bessel function of the second kind, which decays exponentially for large $z$ and is even in $z$. 

We now use $G(\tau)$ to calculate the first order perturbation term $\langle S_1 \rangle$ in Eq.~(\ref{S_pert}) and rescale $\bar \tau= \tau/ b$. 
\begin{equation}
\begin{aligned}
\langle S_1 \rangle 	&= -V_0 \Lambda \int d\tau \langle \cos \big(\phi_<(\tau) + \phi_>(\tau) \big) \rangle \\
				&= -V_0 \Lambda \int d\tau \Big[ \cos \phi_< \langle \cos \phi_> \rangle - \sin \phi_< \langle \sin \phi_> \rangle \Big] \\
				&= -V_0 \Lambda \int d\tau \cos \phi_< \Big[ e^{ -\frac{1}{2} G(0)} \Big] \\
				&= -V_0 \Lambda b^{ 1 -\frac{1}{\alpha}} \int d \bar \tau \cos (\phi_< (\bar \tau)),
\end{aligned}
\end{equation}
where in line 2, $\langle \sin \phi_> \rangle$ vanishes since $\sin \phi$ is odd in $\phi$. 
Therefore, first order perturbation rescales $V_0$ as $V_0(b) = b^{1-\frac{1}{\alpha}}V_0$ and does not rescale $\alpha$. Upon linearizing ${b\approx 1+\delta l}$, we arrive at the following differential flow equation,
\begin{equation}
\frac{\partial V_0(l)}{\partial l} = \Big(1-\frac{1}{\alpha} \Big) V_0(l), \label{O1}
\end{equation}
and observe that $V_0$ is irrelevant and scales to 0 if $\alpha < 1$ and grows if $\alpha > 1$. 

Next, we calculate the second order perturbation term. Denoting $\phi(\tau)$ and $\phi(\tau')$ as $\phi$ and $\phi'$, respectively,
\begin{equation}
\begin{aligned}
\langle S_1^2 \rangle 	&= (V_0 \Lambda)^2 \int d\tau d\tau' \langle \cos \big(\phi_< + \phi_> \big) \cos \big(\phi_<' + \phi_>' \big) \rangle \\ 		
					&= (V_0 \Lambda)^2  \int d\tau d\tau' \frac{e^{-G(0)}}{2}  \Big[ \cos (\phi_< - \phi'_<) e^{G(\tau-\tau')} \\
					&  \qquad  \qquad \qquad + \cos (\phi_< + \phi'_<) e^{-G(\tau-\tau')} \Big],		
\end{aligned}
\end{equation}
and thereby, 
\small
\begin{align}
\langle S_1^2 \rangle - \langle S_1 \rangle^2 =& \frac{(V_0 \Lambda)^2}{2} e^{-G(0)} \int d\tau d\tau'  \label{O2}\\
					\times \Big[ \cos (\phi_< - \phi'_<) &(e^{G(\tau-\tau')}-1) + \cos (\phi_< + \phi'_<) (e^{-G(\tau-\tau')}-1) \Big]. \nonumber
\end{align}
\normalsize
By gradient expansion, the first term $\sim \Big(\frac{\partial \phi}{\partial \tau}\Big)^2 \sim \omega^2$, and is therefore irrelevant by power counting, as with the quadratic term in $S_0$. The second term $\sim \cos(2\phi)$ is analogous to the symmetry breaking field $h_p \cos (p\theta(x))$ of the 2D sine-Gordon Hamiltonian in Eq.~(\ref{HSG}). As the 1D dissipation model and the 2D sine-Gordon model utilize the same renormalization group, their scaling relations can be mapped onto each other. From this mapping, we can then obtain the relevance conditions for the periodic potentials in the 1D QBM model. Specifically, for any term of the form $V_n \cos(n\phi)$ in the 1D QBM action, we make the following correspondence to the relation derived for the roughening transition \cite{Kardar},
\begin{equation}
\frac{d h_p}{d l} = \Bigg( 2-\frac{p^2}{4 \pi J} \Bigg) h_p \quad \Longleftrightarrow \quad \frac{d V_i}{d l} = \Bigg( 1-\frac{n^2}{\alpha} \Bigg)V_i,
\end{equation}
where $2 \rightarrow 1$ in the first term comes from going from two-dimensional real space to one-dimensional time, and $4 \pi J \rightarrow \alpha$ in the second term is found by comparing the two-point correlation functions of the bare actions of both models.

Thus, for the $V_n \cos n\phi$ term to be relevant, we need $\alpha > n^2$. However, for the perturbation relations in $V_0$ to be valid, we need $\alpha < 1$, as shown by the first order relation in Eq.~(\ref{O1}). Therefore, $\alpha > n^2$ cannot be satisfied and the $V_n \cos(n\phi)$ term \big($\cos(2\phi)$ in this specific case\big) is not relevant. Thus, to second order, neither $V_0$ nor $\alpha$ is rescaled. 

In fact, as argued in Ref.~\onlinecite{Fisher}, we do not expect $\alpha$ to be rescaled to any order in $V_0$. As defined, $\alpha$ depends only on $\eta$ and $x_0$. As we have pointed out before, $x_0$ is a system constant that does not change under renormalization. On the other hand, $\eta$ is the coefficient of $|\omega||\tilde \phi(\omega)|^2$, a term that cannot be generated by derivatives of $\phi(\tau)$ and is thus nonlocal, as explicitly apparent in its Fourier transform: $\big(\frac{x(\tau) - x(\tau')}{\tau - \tau'} \big)^2$. Therefore, since the perturbative RG procedure is only capable of generating local interactions, $\eta$ and thus $\alpha$ remain unaltered up to all orders in $V_0$. This is markedly different from the behavior of the 2D sine-Gordon model. Even though both utilize the 2D Coulomb gas renormalization group technology, the $~q^2|\tilde \theta(q)|^2$ term in the 2D sine-Gordon model is local and the temperature coefficient $J \sim T^{-1}$ is rescaled under renormalization. 

In the resulting flow diagram, this means that the line of the phase transition is vertical, as shown in Fig.~\ref{schm}. Since the system is self-dual in the weak and strong coupling limit, the same analysis can be applied in the $V\rightarrow \infty$ regime. As such, Refs.~\onlinecite{Fisher, Schmid} have used self-duality and RG analysis to derive the flow lines in the strong coupling limit and the perturbative weak coupling limit. 
For low enough friction $\alpha <1$, $V$ scales to 0. In other words, at zero temperature (infinite imaginary timescales) where there is no finite energy scale to cut off the renormalization group flow, the system's behavior is described by the fixed line $V=0$ and the particle is delocalized. On the other hand, for high enough friction $\alpha >1$, $V$ scales to infinity, and the particle is localized.

\section{Rectangular and square lattice potentials} \label{2D_Uncoupled}

Starting in this section, we extend RG analysis of a particle in a 1D periodic potential to that in a 2D periodic potential. 

A particle in two dimensions has two orthogonal modes: $x(\tau)$ and $y(\tau)$. For our analysis, we assume that the friction coefficient itself, $\eta$, is isotropic. Then, given lattice parameters $x_0$ and $y_0$, we rescale and denote our dimensionless parameters accordingly, 
\begin{equation}\label{2D_par}
\begin{aligned}
\alpha_x &=&& \frac{\eta x_0^2}{2\pi }, 	\qquad& \alpha_y &=&&\frac{\eta y_0^2}{2\pi },\\
V_x &=&& \frac{V^x}{\Lambda}, 			\qquad& V_y &= &&\frac{V^y}{\Lambda}, \\
\phi_x(\tau) &=&& \frac{2 \pi x(\tau)}{x_0}, 	 \qquad& \phi_y(\tau) &=&&\frac{2 \pi y(\tau)}{y_0}.
\end{aligned}
\end{equation}
The bare action is then
\begin{equation}
S_0 = \frac{1}{2} \int \frac{d\omega}{2\pi} \Big[ \tilde K_x(\omega) |\tilde \phi_x(\omega)|^2 +  \tilde K_y(\omega) |\tilde \phi_y(\omega)|^2 \Big],
\end{equation}
where $\tilde K_i(\omega) = \frac{\alpha_i}{2 \pi} |\omega| + \frac{\omega^2}{\Lambda}$ is as defined previously for each single mode in a 1D periodic potential. 

We note that we now have two separate dimensionless parameters $\alpha_x$ and $\alpha_y$ that each, when considered by itself, only affects the localization behavior of a single mode. They depend on the same friction coefficient $\eta$ but different lattice parameters $x_0$ and $y_0$, as seen in Eq.~(\ref{2D_par}). Thus, we expect that anisotropy in the lattice geometry will affect the transition points of a system. In particular, there may exist regions in parameter space where the two modes have different localization behaviors. This is indeed what we find in the calculations that follow. 

A rectangular lattice potential, with adjacent minima in the $x$ and $y$-directions separated by $x_0$ and $y_0$, respectively, is given by
\begin{equation} \label{rect_hamiltonian}
\begin{aligned}
S_1 	&= -\int d\tau \left[ V^x \cos\Big( \frac{2\pi x(\tau)}{x_0}\Big) + V^y \cos\Big( \frac{2\pi y(\tau)}{y_0}\Big) \right]\\
	&= S_1^x(x) +  S_1^y(y).
\end{aligned}
\end{equation}
After rewriting the potentials in terms of the dimensionless variables and parameters defined in Eq.~(\ref{2D_par}), we can rewrite the entire action $S=S_0 + S_1$ as separate functions of $\phi_x(\tau)$ and $\phi_y(\tau)$:
\begin{equation} 
\begin{aligned}
S 	&= S^x(\phi_x) + S^y(\phi_y),
\end{aligned}
\end{equation}
where,
\small
\begin{equation}
S^i(\phi_i) = \frac{1}{2} \int \frac{d\omega}{2\pi} \tilde K_i (\omega) |\tilde \phi_i(\omega)|^2 - V_i \Lambda \int d\tau \cos \big( \phi_i(\tau) \big).
\end{equation}
\normalsize
Now we can solve for each mode independently as we did for the particle in one dimension. 

Performing RG analysis for each mode gives us two sets of separate conditions regarding $\alpha_x$ and $\alpha_y$, which impose conditions on the relevance of $V_x$ and $V_y$, respectively. Since $\alpha_x$ and $\alpha_y$ are rescaled versions of the same phenomenological parameter $\eta$, we obtain the following conditions,
\begin{equation}
\frac{\eta}{2 \pi } \begin{cases}
		\begin{cases}
              	< \frac{1}{x_0^2} , \text{	$V_x$ is irrelevant}\\
		> \frac{1}{x_0^2} , \text{	$V_x$ is relevant}\\
		\end{cases}
		\\
		\\
		\begin{cases}
		< \frac{1}{y_0^2} , \text{	$V_y$ is irrelevant}\\
		> \frac{1}{y_0^2} , \text{	$V_y$ is relevant.}\\
		\end{cases}
            \end{cases}
 \end{equation}

\begin{figure}[h]
\includegraphics[width=1\columnwidth]{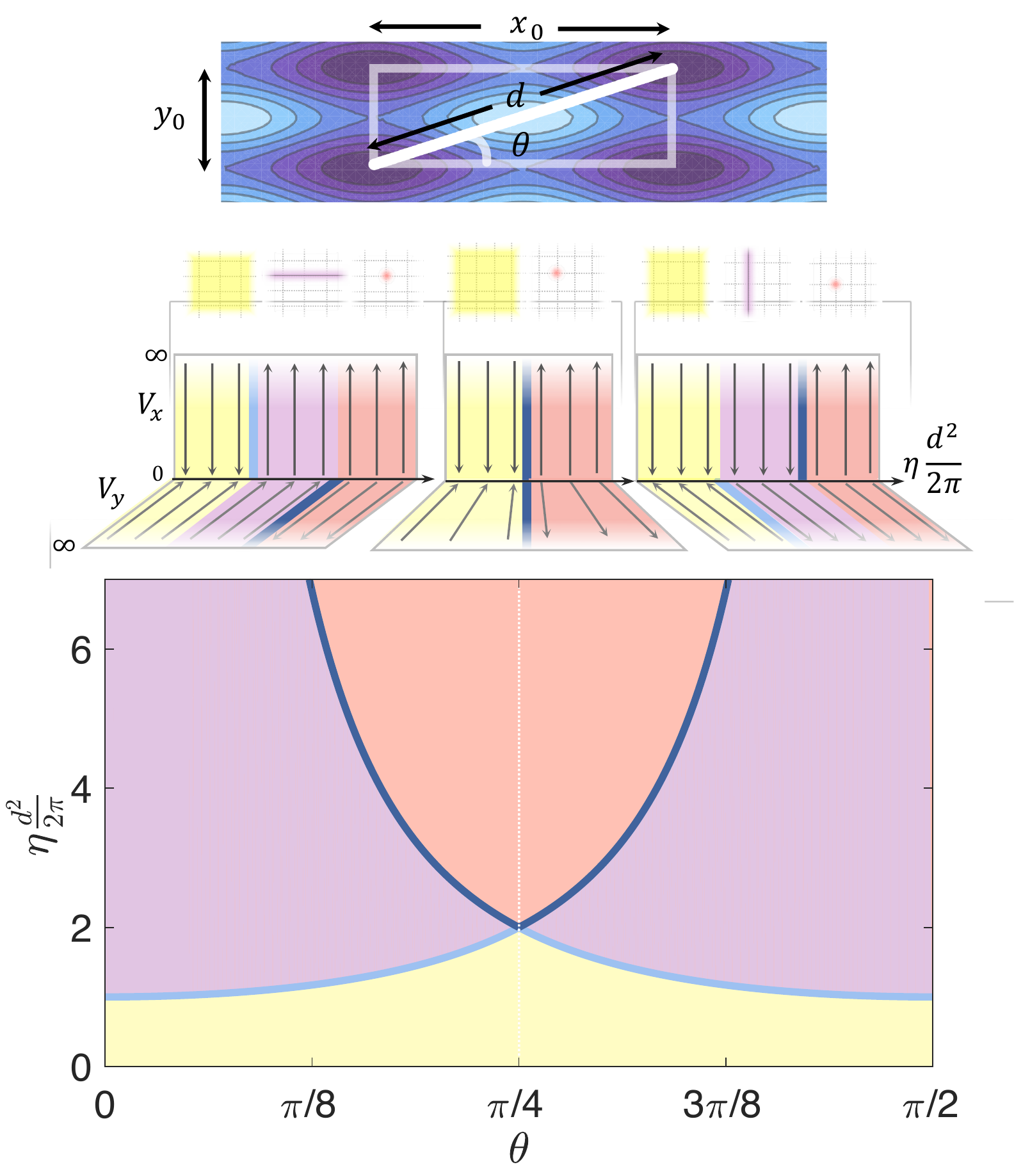}
\caption{\label{rect} Flow diagrams for a particle in a dissipative environment coupled to a rectangular potential are drawn for three different ranges of $\theta$, as denoted in the top figure: $0<\theta < \pi/4$, $\theta = \pi/4$, and $ \pi/4 < \theta < \pi/2$. The illustrations above each flow diagram are schematics that indicate the phases exhibited by the system for a given value of the friction coefficient $\eta$. The phase diagram of the particle are plotted as a function of $\theta$ for fixed $d$. For a non-square potential, the particle exhibit a phase of 1D line-localization (purple) before passing from the delocalized phase (yellow) into the point-localized phase (red). The dotted line at $\theta = \pi/4$ corresponds to the square lattice potential.} 
\end{figure}

\begin{center}
\begin{table}
\begin{tabular} {| c | c | c | c|}
\hline
	$\theta$ range	& phases			&	$\frac{d^2}{2 \pi } \cdot \eta(\theta) $ \\
\hline
\hline
\multirow{4}{*}{$0<\theta< \frac{\pi}{4}$} 	& \multirow{2}{*}{delocalized} & \multirow{3}{*}{--------- $\sec^2(\theta)$ ---------}\\
		&  \multirow{2}{*}{1D $x$ loc.} 	&  \multirow{3}{*}{--------- $ \csc^2(\theta)$ ---------} \\
		&   \multirow{2}{*}{2D loc.} 			&  \\
		&&\\
\hline
\multirow{3}{*}{$ \theta = \frac{\pi}{4} $} 		& \multirow{2}{*}{delocalized} & \multirow{3}{*}{--------------$1$--------------}  \\
		& \multirow{2}{*}{2D loc.} 			& \\
		&& \\
\hline
\multirow{4}{*}{$\frac{\pi}{4} < \theta<\frac{\pi}{2}$} 	& \multirow{2}{*}{delocalized} & \multirow{3}{*}{--------- $\csc^2(\theta)$ ---------}\\
		&  \multirow{2}{*}{1D $y$ loc. } 	&  \multirow{3}{*}{--------- $ \sec^2(\theta)$ ---------} \\
		&   \multirow{2}{*}{2D loc.} 			&  \\
		&&\\
\hline
\end{tabular}
{\caption {\label{rect_trans}} The possible phases exhibited by a 2D QBM model in a rectangular lattice potential, and the corresponding phase boundaries as a function of the lattice parameters, where $d$ and $\theta = \tan^{-1}(y_0/x_0)$ are as labeled in Fig.~\ref{rect}. }
\end{table}
\end{center} 

For a square lattice potential where $x_0=y_0$, the two sets of conditions are identical, and we obtain the same zero temperature phase diagram as that of the one dimensional case, with a phase transition from the delocalized phase to the localized phase at $\alpha_x = \alpha_y = 1$. 

For rectangular lattice potentials where $x_0 \neq y_0$, we get that the relevance of $V_x$ and $V_y$ depend differently on  $\eta$. More concretely, denoting $\eta_i$ as the minimum value of $\eta$ that leads to localization of the particle in the $i$-direction, and letting $x_0 > y_0$, we see that $\eta_x < \eta_y$. Therefore, we obtain the following picture: in the direction where the adjacent minima are farther apart, it is harder for the particle to tunnel and so it requires less friction to ``get stuck.'' The possible zero temperature phase diagrams, in a plane with fixed coordinates, are shown in Fig.~\ref{rect}. Denoting $d^2 \equiv x_0^2+y_0^2$ and $\tan \theta \equiv y_0/x_0$, the phase boundaries are summarized in Tab.~\ref{rect_trans}.


\section{Centered rectangular and triangular (hexagonal) lattice potentials} \label{rhSec}

We now consider centered rectangular potentials obtained by the multiplication of periodic functions, 
\begin{equation} \label{rhPot}
\begin{aligned}
S_0 &= \frac{1}{2} \int \frac{d\omega}{2\pi} \Big[ \tilde K_x(\omega) |\tilde \phi_x(\omega)|^2 +  \tilde K_y(\omega) |\tilde \phi_y(\omega)|^2 \Big] \\
S_1 &= V_{xy} \Lambda \int d\tau \cos\Big(\phi_x(\tau)\Big) \cos\Big( \phi_y(\tau)\Big),
\end{aligned}
\end{equation}
where $x_0$ and $y_0$ are again the distance between the adjacent potential minima in the $x$ and $y$-direction, respectively, and where $\tilde K_i(\omega) = \frac{\alpha_i}{2 \pi} |\omega| + \frac{\omega^2}{\Lambda}$ is as defined in the previous section. In contrast to the rectangular lattice, the rows of potential wells are staggered (see schematic in Fig.~\ref{rh_pd}). Physically, we expect the anisotropy of the lattice parameters to affect the phase transition boundaries but must more carefully examine the effects of $V_{xy}$, which now couples the two orthogonal modes $x(\tau)$ and $y(\tau)$. 

We cannot perform coordinate transformation to decouple modes in $S_1$ without inducing mode-mixing in $S_0$. That is, the following transformation $\phi_x' = \phi_x + \phi_y$, $\phi_x' = \phi_x - \phi_y$ gives us decoupled $S_1$,
\begin{equation}
S_1 = \frac{V_{xy} \Lambda}{2} \int d\tau \cos(\phi_x'(\tau)) + \cos(\phi_y'(\tau)),
\end{equation}
but rearranges $S_0$ as
\begin{equation}
\begin{aligned}
S_0 &= \int \frac{d\omega}{2\pi}  \frac{1}{4\pi} \Bigg[(\alpha_x + \alpha_y)\bigg(|\tilde \phi_x'(\omega)|^2 + |\tilde \phi_y'(\omega)|^2 \bigg) \\
	& \qquad   +  (\alpha_x - \alpha_y)\bigg(\tilde \phi_x'(\omega) \tilde \phi_y'(-\omega)+\tilde \phi_x'(-\omega) \tilde \phi_y'(\omega) \bigg) \Bigg],
\end{aligned}
\end{equation}
where the mode-mixing term only disappears when ${x_0 = y_0}$, corresponding to the case of a square lattice with inter-minima spacing of $x_0/\sqrt{2}$. Therefore, we perform perturbative RG to calculate the effect of $S_1$ on the system parameters. 

\subsection{Perturbative RG analysis} \label{2Dnaive}

Intermediate steps of the calculation are shown in Appendix \ref{apA}. We will summarize the results in this section. 
Using Eq.~(\ref{gtau}), we identify the two-point correlation functions of the bare action upon integrating over the fast modes $G_i(\tau)  \equiv \langle \phi_{i,>} (\tau) \phi_{i,>} (0) \rangle$,
\begin{equation}
\begin{aligned}
G_i(\tau) 	&= \frac{2}{\alpha_i} Y_0(\Lambda \tau /b) \quad \text{for } \Lambda\tau \gg 1, \\
G_i(0) &= \frac{2}{\alpha_i}\ln(b),
\end{aligned}
\end{equation}
where $i$ = $x$, $y$ and $\langle \phi_{x,>} (\tau) \phi_{y,>} (0) \rangle = 0$. We then obtain that the first order perturbation $\langle S_1 \rangle$ leads to the following scaling relation,
\begin{equation}
\frac{\partial V_{xy}(l)}{\delta l} = \Bigg[1-\Big(\frac{1}{\alpha_x}+\frac{1}{\alpha_y}\Big) \Bigg] V_{xy}(l).
\end{equation}
Thus, $V_{xy}$ scales to 0 when $ \frac{\eta}{2 \pi } < \big( \frac{1}{x_0^2}+\frac{1}{y_0^2} \big)$. We check that when $x_0=y_0$, the imposed condition becomes identical to that of a square lattice with spacing $x_0/\sqrt{2}$. 
Physically, we expect the anisotropy of the potential to reflect in the system in some way, which motivates us to seek higher order effects. 
We find that upon calculating the second order effect $ \langle S_1^2 \rangle$, the following terms are generated,
\begin{equation}
\begin{aligned}
V_{xy}^{(2)} \cos(2\phi_x) \cos(2 \phi_y) + V_x \cos(2\phi_x) + V_y \cos(2\phi_y), 
\end{aligned}
\end{equation}
where, using results (\ref{exp1}) and (\ref{exp2}), the coefficients are, to this order in $V_{xy}$,
\begin{equation}
\begin{aligned}
V_{xy}^{(2)} &= \frac{(V_{xy}\Lambda)^2}{4}e^{-(G_x(0)+G_y(0))} \\
		&  \qquad \qquad \times \Big\{\cosh \Big[ G_x(\tau-\tau')+G_y(\tau-\tau')\Big]-1 \Big\} \\
V_x &= V_y = \frac{(V_{xy}\Lambda)^2}{4}e^{-(G_x(0)+G_y(0))} \\
		&  \qquad \qquad \times \Big\{\cosh \Big[ G_x(\tau-\tau')-G_y(\tau-\tau')\Big]-1 \Big\}. \\
\end{aligned}
\end{equation}

Now, we analyze the relevance of each of the generated terms. We immediately see that $V_{xy}^{(2)}\cos(2\phi_x)\cos(2\phi_y)$ is irrelevant, and comment that this is true for any term of the form $V_{xy}^{(n)} \cos(n\phi_x)\cos(n\phi_y)$ where $n>1$. In fact, we see that, generally, a term of the form $\cos(n_x\phi_x)\cos(n_y\phi_y)$ effectively rescales $x_0 \rightarrow x_0/n_x$ and $y_0 \rightarrow y_0/n_y$ and is relevant in the presence of our current potential term $V_{xy}\cos(\phi_x)\cos(\phi_y)$ if the following condition is satisfied,
\begin{equation} \label{clock}
\frac{n_x^2}{\alpha_x} + \frac{n_y^2}{\alpha_y} < \frac{1}{\alpha_x} + \frac{1}{\alpha_y} .
\end{equation}
Thus, denoting $\tan(\theta) \equiv y_0/x_0$, we arrive at the conditions that in the perturbative regimes of both $V_x$ and $V_{xy}$, $V_x\cos(2\phi_x)$ ($n_x = 2, n_y = 0$) is relevant when $\theta < \pi/6$, and $V_y\cos(2\phi_y)$ ($n_x = 0, n_y = 2$) is relevant when $\theta > \pi/3$. 
Here, we ponder if additional, different, terms can be generated by higher order perturbations. However, we see that the only terms that have not yet been generated and which have possible lattice configurations that can satisfy condition (\ref{clock}) are those of ($n_x>2, n_y = 0$) and ($n_x=0, n_y > 2$). These terms are irrelevant in the presence of the $\cos(2\phi_i)$ terms that have already been generated. 
Now, denoting $d$ and $\theta$ as in the top left of Fig.~\ref{rh_pd}, we note that unlike that for the rectangular potential, there exists a finite range in $\theta$ where the system transitions directly from the diffuse phase to the point localized phase, skipping over the line localized phase.

\subsection{Behavior of a mixing potential under strong coupling in one direction} \label{2Dredo}

In Sec.~\ref{2Dnaive}, we considered the regime in which all potential coupling terms are perturbative. In the single dimensional line localization phase, however, the parameter which dictates the direction of localization is no longer in this perturbative regime. We must therefore consider these situations more carefully. 

For completely decoupled potential terms, as written for the rectangular potential in Eq.~(\ref{rect_hamiltonian}), the relevance of $V_i$ is independent of the value of $V_j$ for $i\neq j$. 

Here, we examine the relevance of the perturbative coupling term $V_{xy}$ in the regime of strong coupling in the $x$-direction, where $V_x \rightarrow \infty$, such that 
\small
\begin{equation}
\begin{aligned}
S_0 &= \frac{1}{2} \int \frac{d\omega}{2\pi} \Bigg[\frac{\alpha_x}{2 \pi} |\omega| + \frac{\omega^2}{\Lambda} \Bigg] |\tilde \phi_x(\omega)|^2 + \Bigg[\frac{\alpha_y}{2 \pi} |\omega| + \frac{\omega^2}{\Lambda} \Bigg] |\tilde \phi_y(\omega)|^2 \\
&\qquad  \qquad - V_x \Lambda \int d\tau \cos \big( \phi_x(\tau) \big) \\
S_1 &= - V_{xy} \Lambda \int d\tau \cos \big( \phi_x(\tau) \big) \cos \big( \phi_y(\tau) \big).
\end{aligned}
\end{equation}
\normalsize
We show below, using two different methods, that in the strong coupling regime of $V_x$, the scaling of the mixing potential becomes identical to that of a unidirectional $V_y$ potential. 

\subsubsection{Method 1 - RG analysis of cosine expansion} \label{method1}
 
We examine the two-point correlation function $G_x(\tau)$ in the regime of $V_x \rightarrow \infty$ by expanding the cosine potential term. 
\small
\begin{equation}
\begin{aligned}
	& -\int d\tau V_{x}\Lambda \cos (\phi(\tau)) \\
=	& -\int d\tau V_{x}\Lambda \sum_n \frac{(-1)^n \phi^{2n}}{(2n)!} \\
=	& -V_{x}\Lambda \sum_n \frac{(-1)^n}{(2n)!} \int \prod_{i=1}^{2n-1} \frac{d\omega_{i}}{2\pi} \tilde \phi(\omega_i) \times \tilde \phi(-\omega_1-\ldots-\omega_{2n-1})
\end{aligned}
\end{equation}
\normalsize
In the power expansion, each $d\omega_i$ gets scaled under renormalization as $\omega = \omega b^{-1}$. Therefore, each term in the expansion is scaled by $b^{1-2n}$, and the term lowest order in $d\omega \phi(\omega)$ dominates. The $n=0$ term is a constant shift that we ignore. Therefore we have left over from the strong coupling term, 
\begin{equation}
V_x\Lambda \frac{1}{2} \int d\omega |\tilde \phi(\omega)|^2,
\end{equation}
which gives the new $\tilde K(\omega)$ previously described in Eq.~(\ref{scaledH}) as,
\begin{equation}
\tilde K(\omega) = V_x\Lambda+\frac{\alpha}{2\pi}|\omega|+\frac{\omega^2}{\Lambda}.
\end{equation}
In Sec.~\ref{1DRG}, we argued the quadratic kinetic energy term to be irrelevant by power counting in the presence of the lower order friction term. Here, we see that both the friction term and the quadratic term are irrelevant by power counting in the presence of $V_x\Lambda$. It is important to emphasize again that we are able to say this because the periodic constraint of the system prevents $\phi$ itself from rescaling under renormalization. 

Then, the two-point correlation $G_x(\tau)$ is given by the expression,
\begin{equation}
G_x(\tau) \equiv \langle \phi_{x,>}(\tau) \phi_{x,>}(0) \rangle = \frac{1}{V_x \Lambda} \int_{\Lambda/b}^\Lambda \frac{d\omega}{2\pi} e^{i \omega \tau}W\Big(\frac{\omega}{\Lambda/b}\Big),
\end{equation}
where $W(x)$ is the smoothing function used in Sec.~\ref{1DRG}. 
From the above we obtain 
\begin{equation}
G_x(0) = \frac{1}{V_x b} \frac{\sqrt{1+b^2}-\sqrt{2}}{4\pi},
\end{equation}
and letting $b\approx1+\delta l$,
\begin{equation}
\begin{aligned}
G_x(0)	& = \frac{\gamma}{V_x}\Big( 1-\rho \delta l \Big)\\
		& \approx \frac{\gamma}{V_x} b^{-\rho}, 
\end{aligned}
\end{equation}
where $\gamma = \frac{\sqrt{2}-1}{4 \pi}$ and $\rho = \frac{1}{\sqrt{2}}$ are positive numerical constants. 

Using this result, we calculate the first order perturbation term,
\begin{equation}\label{H1_att1}
\begin{aligned}
\langle S_1 \rangle 	& = \exp\Bigg[-\frac{1}{2}\bigg(G_x(0)+G_y(0)\bigg)\Bigg] \\
				& = b^{-\frac{1}{\alpha_y}}\exp\Bigg[-\frac{\gamma}{V_x}b^{-\rho}\Bigg].  
\end{aligned}
\end{equation}
In the regime that we are concerned with, where $V_x > \alpha_y$, the exponential term in expression (\ref{H1_att1}) decays to 1 fast enough to be negligible and the product is dominated by $b^{-1/\alpha_y}$. Using the Heavyside function instead of the smoothing function $W(x)$ to calculate $G_x(0)$ gives the same qualitative results here. We thereby obtain that in the strong coupling regime of $V_x \rightarrow \infty$, $V_{xy}$, the potential term mixing directions $x$ and $y$, scales in a way that is identical to that of an uncoupled potential in the $y$-direction $V_y$. Physically, this says that in the $V_x \rightarrow \infty$ regime, the fluctuations are dominated by those in the $y$-direction, which then determine the relevance of the mixing potential $V_{xy}$. 

\subsubsection{Method 2 - RG analysis of strong coupling expansion}

We also study the strong coupling regime using the instanton gas picture, as reviewed in Sec.~\ref{instanton}. We first comment that the effect of RG on an instanton gas configuration $\bar \phi (\tau)$, from Eqs.~(\ref{single_inst}) and (\ref{inst_gas}), seems to be the following: Since $\bar \phi$ itself is not renormalized while $\tau$ is rescaled, $\bar \phi$ can be squeezed/stretched in time space without affecting the underlying symmetry of the periodic potential. Simplistically, coarse-graining by removing high frequency components leads to instanton gas configurations where neighboring instantons are at least $b\omega_0^{-1}$ away from each other. Rescaling $\bar \tau=\tau/b$ then squeezes each instanton/anti-instanton in time space. 

To calculate the two-point correlation function $G_x(\tau)$, we proceed as follows. First, the energy of a neutral configuration consisting of $N$ instantons and $N$ anti-instantons is, following calculations from Sec.~\ref{instanton},
\begin{equation}
H[\phi_{2N}] = 2Ns-\alpha \sum_{j\neq k} e_j e_k \ln \Big(\omega_0 (\tau_j-\tau_k)\Big), 
\end{equation}
where $s$ is the self-interacting energy of each instanton/anti-instanton, and core-range interactions are neglected since under coarse-graining, the surviving configurations are such that $\omega_0 \Delta \tau \geq b > 1$. 
In light of our analysis of centered rectangular potentials, the one-dimensional potential that emerges from perturbative RG is of the form $V_x \cos(2\phi_x)$. Therefore, in solving for the instanton solution in Eq.~(\ref{instantonDiff}), we rescale $\phi_x \rightarrow 2 \phi_x$, modifying the above equation to be,
\begin{equation} \label{instantonH}
H[\phi_{2N}] = 2Ns-\frac{\alpha_x}{4} \sum_{j\neq k} e_j e_k \ln \Big(\frac{\omega_0}{\sqrt{2}} (\tau_j-\tau_k)\Big). 
\end{equation}

Then, including the effects of fluctuations in the prefactor, we have that
\begin{equation} 
\begin{aligned}
	& e^{-H[\phi_{2N}]} \\
= 	& \Bigg[\frac{\big(\frac{\omega_0}{\sqrt{2}}\sqrt{\frac{2 s}{\pi}}e^{-s}\big)^N}{N!}\Bigg]^2 \prod^{2N}_{j\neq k} \Big(\frac{\omega_0}{\sqrt{2}} (\tau_j-\tau_k)\Big)^{\frac{\alpha_x}{4} e_j e_k }. 
\end{aligned}
\end{equation}
Assuming only dilute instanton gas configurations, we again neglect core-range interactions, and the partition function $Z$ becomes,
\begin{equation}
\begin{aligned}
Z =	&  \int d\tau_1\ldots d\tau_{2N} e^{-H[\phi_{2N}]} \\
   =& \sum_N \Bigg[\frac{\big(\frac{\omega_0}{\sqrt{2}}\sqrt{\frac{2 s}{\pi}}e^{-s}\big)^N}{N!}\Bigg]^2 \sum_{\{\tau_i, e_i\}} \int_{-\infty}^{\infty} d\tau_1 \\
	&  \qquad \times \int_{\omega_0^{-1}}^{\infty} d\bar \tau_2 \ldots d\bar \tau_{2N} \prod_{j\neq k}^{2N} \Big(\frac{\omega_0}{\sqrt{2}} (\tau_j-\tau_k)\Big)^{\frac{\alpha_x}{4} e_j e_k },
\end{aligned}
\end{equation}
where in the second step, we have defined ${\bar \tau_i = \tau_i - \tau_{i-1}}$, and $\sum_{\{\tau_i, e_i\}}$ sums over all possible assignments of ${\{ e_i = \pm 1 \}}$ for a given neutral instanton gas density $\{\tau_i\}$ (i.e. all distinct sequences of instantons/anti-instantons). 

We wish to examine the two-point correlation function $G_x(\tau)$ by averaging over all instanton gas configurations in the presence of the logarithmic potentials that arise from interactions with the dissipative environment. Looking specifically at $G_x(0)$, we have,
\begin{equation}
\begin{aligned}
	& \langle \phi_x(0) \phi_x(0) \rangle \\
=	& \frac{1}{Z} \sum_N \Bigg[\frac{\big(\frac{\omega_0}{\sqrt{2}}\sqrt{\frac{2 s}{\pi}}e^{-s}\big)^N}{N!}\Bigg]^2 \sum_{\{\tau_i, e_i\}} \int d\tau_1\ldots d\tau_{2N} \\
	&  \times \prod_{j\neq k}^{2N} \Big(\frac{\omega_0}{\sqrt{2}} (\tau_j-\tau_k)\Big)^{\frac{\alpha_x}{4} e_j e_k } \sum_{m, n=1}^{2N} e_m e_n f(-\tau_m) f(-\tau_n),
\end{aligned}
\end{equation}
where $\pm f(\tau) = \pm 4 \arctan(e^{\omega_0 \tau})$ is the solution corresponding to a single instanton/anti-instanton from Eq.~(\ref{single_inst}). 
After coarse graining, rescaling, and imposing the similarity condition $Z=Z'$, we have that 
\small
\begin{equation} \label{RGInst}
\begin{aligned}
	& \langle \phi_x(0) \phi_x(0) \rangle \\
=	& \frac{1}{Z}	\sum_N \Bigg[\frac{\big(\frac{\omega_0}{\sqrt{2}}\sqrt{\frac{2 s}{\pi}}e^{-s}\big)^N}{N!}\Bigg]^2 \sum_{\{\tau_i, e_i\}}  \int_{-\infty}^{\infty} d\tau_1\int_{\omega_0^{-1}}^{\infty} d\bar \tau_2 \ldots d\bar \tau_{2N}  \\
	&  \times \prod_{j\neq k}^{2N} \Big(\frac{\omega_0}{\sqrt{2}} (\tau_j-\tau_k)\Big)^{\frac{\alpha_x}{4} e_j e_k } \sum_{m, n=1}^{2N} e_m e_n f(-b^2 \tau_m) f(-b^2 \tau_n). 
\end{aligned}
\end{equation}
\normalsize
Details of the calculation leading up to the final result may be found in Appendix \ref{apB}. Upon examining each term in the $N$-summation individually, substituting in the expression for $f(\tau)$ from Eq.~(\ref{single_inst}), and taking into account the invariance of the absolute position of an instanton gas configuration in time space, we arrive at the following result,
\begin{equation}
\lim_{b\rightarrow \infty}  G_x(0) = \pi^2.
\end{equation}

Since $G_x(0)$ approaches a finite constant as $b$ approaches infinity, the behavior of $be^{-(G_x(0)+G_y(0))}$, which scales $V_{xy}$ to first order, is dominated by $b^{1-\frac{1}{\alpha_y}}$. Therefore, we reach the same conclusion as we did using method 1 in Sec.~\ref{method1}: in the regime of $V_x \rightarrow \infty$, the coupled potential term $V_{xy}\cos(\phi_x)\cos(\phi_y)$ scales as a single-dimensional potential term $V_y\cos(\phi_y)$. 
\newline

We now put these results back in the context of the centered rectangular potential (Eq.~(\ref{rhPot})). In Sec.~\ref{2Dnaive}, we concluded that in the regime where $V_{xy}$ is perturbative, if $\theta<\pi/6$, a relevant single-dimensional potential term emerges from second order perturbative RG. Then, it follows from this section that as $V_x \rightarrow \infty$, $V_{xy}$ stays irrelevant if the following condition is satisfied,
\begin{equation}
\frac{1}{\alpha_y} < \frac{2^2}{\alpha_x}, 
\end{equation}
or 
\begin{equation}
\theta<\tan^{-1}\Big(\frac{1}{2}\Big) \approx 26.6^\text{o}.
\end{equation}

\begin{figure*}[tb]
\includegraphics[width=\textwidth]{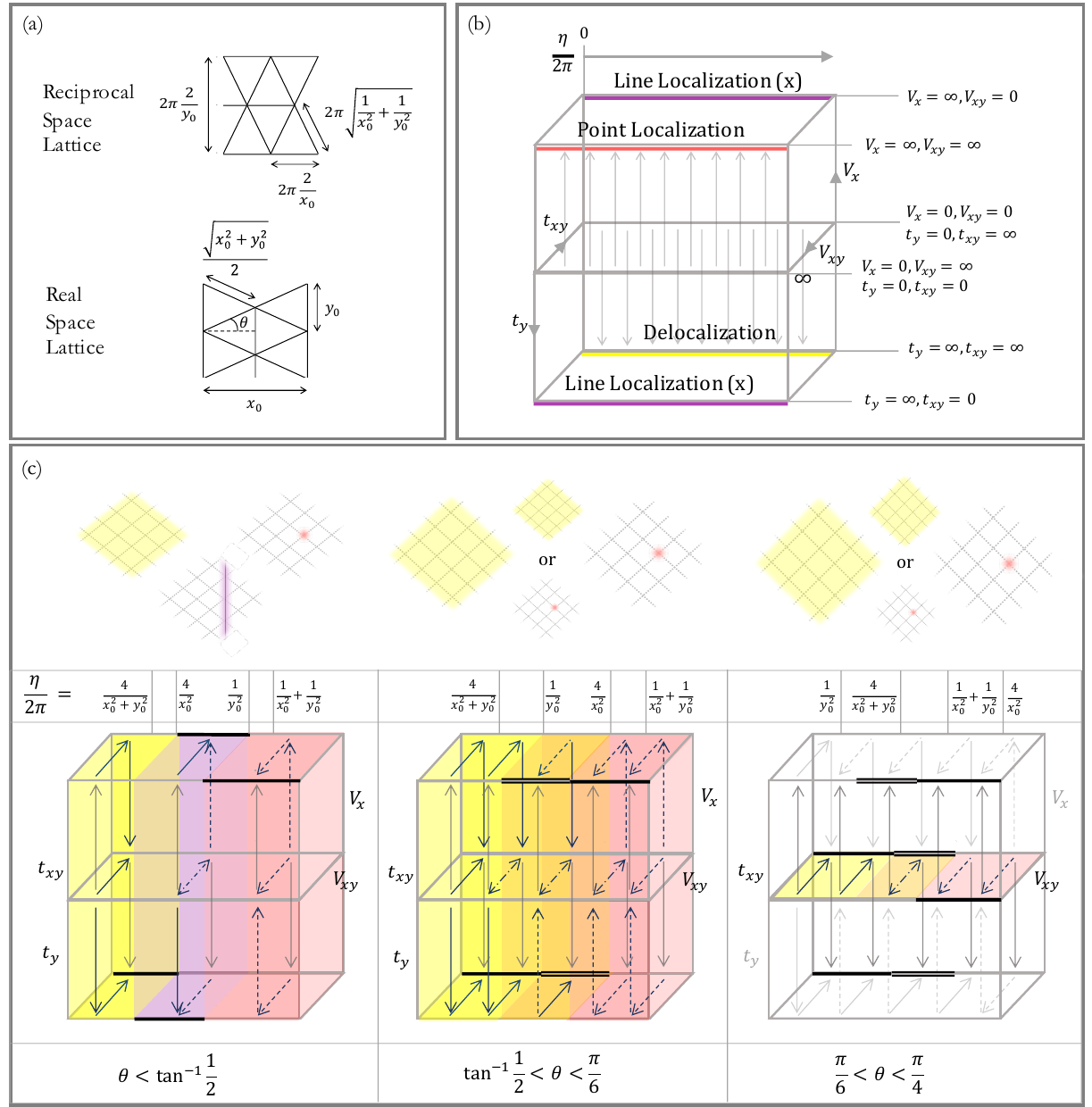}
\caption{\label{rh_flow} (a): Schematic of a general centered rectangular potential for reference and notational clarification. (b): Legend for how to interpret the flow diagrams; the edges of the rectangular prism represent different localization phases. (c): Flow diagrams for three different $\theta$ ranges of the phase diagrams. The illustrations above each flow diagram are schematics which indicate the phases the system exhibits for a given value of the friction coefficient $\eta$. The black lines indicate the stable fixed points, with single-stroke indicating that it is the only stable fixed point for that value of $\eta$ and double-stroke indicating that it is one of two stable fixed points for that value of $\eta$. For the right-most flow diagram, the generated parameters ($V_x$ and $t_y$) are irrelevant.} 
\end{figure*}

\begin{table*}[t]
    \begin{tabular}{| l | c | c |}
    \hline
    ~                                                                                                          		& Real Space 		& Reciprocal Space \\ \hline
    Given parameter ($P$)                                                                                    	& $V_{xy}$          	& $t_{xy}$                \\ \hline
    Generated parameter at 2nd order perturbation ($P_{\rm gen} $)                                                             & $V_{x}$          	& $t_y$                \\ \hline
    Period of generated potential                                                                        	 & $\frac{x_0}{2}$		& $\frac{1}{y_0}$                \\ \hline
    $P$ relevant in the weak coupling regime of $P_{\rm gen} $	     			 & $\frac{1}{\alpha_x} + \frac{1}{\alpha_y} < 1 \Leftrightarrow \frac{\eta}{2 \pi} < \frac{1}{x_0^2} + \frac{1}{y_0^2}$	& $\bar \alpha_x + \bar \alpha_y <1 \Leftrightarrow \frac{2 \pi }{\eta} > \frac{x_0^2 + y_0^2}{4}$                \\ \hline
        $P$ relevant in the strong coupling regime of $P_{\rm gen} $    			& $\frac{1}{\alpha_y} < 1 \Leftrightarrow \frac{\eta}{2 \pi } > \frac{1}{y_0^2}$ & $\bar \alpha_x < 1 \Leftrightarrow \frac{2 \pi }{\eta} < \frac{x_0^2}{4}$                \\ \hline
   $P_{\rm gen} $ relevant in the weak coupling regime of $P$   				& $ \frac{4}{x_0^2} < \frac{\eta}{2 \pi } < \frac{1}{x_0^2} + \frac{1}{y_0^2}$ (need $\theta < \frac{\pi}{6}$)  & $  y_0^2 < \frac{2 \pi }{\eta} < \frac{x_0^2 + y_0^2}{4}$ (need $\theta < \frac{\pi}{6}$)                \\ \hline
    $P_{\rm gen} $ relevant in the strong coupling regime of $P$ 			& $V_{xy} = \infty \rightarrow V_x = \infty$ for all $\eta$ 	& $t_{xy} = \infty \rightarrow t_y = \infty$ for all $\eta$                \\ \hline
    \end{tabular}
    \caption {\label{dualTable} Conditions of relevance for the given potential and tunneling amplitude parameters and those generated under perturbative RG. These conditions are used to construct the flow diagrams in Fig.~\ref{rh_flow}.}
\end{table*}

\subsection{Analysis of the dual reciprocal space lattice}

\begin{figure}[tb]
\includegraphics[width=0.95\columnwidth]{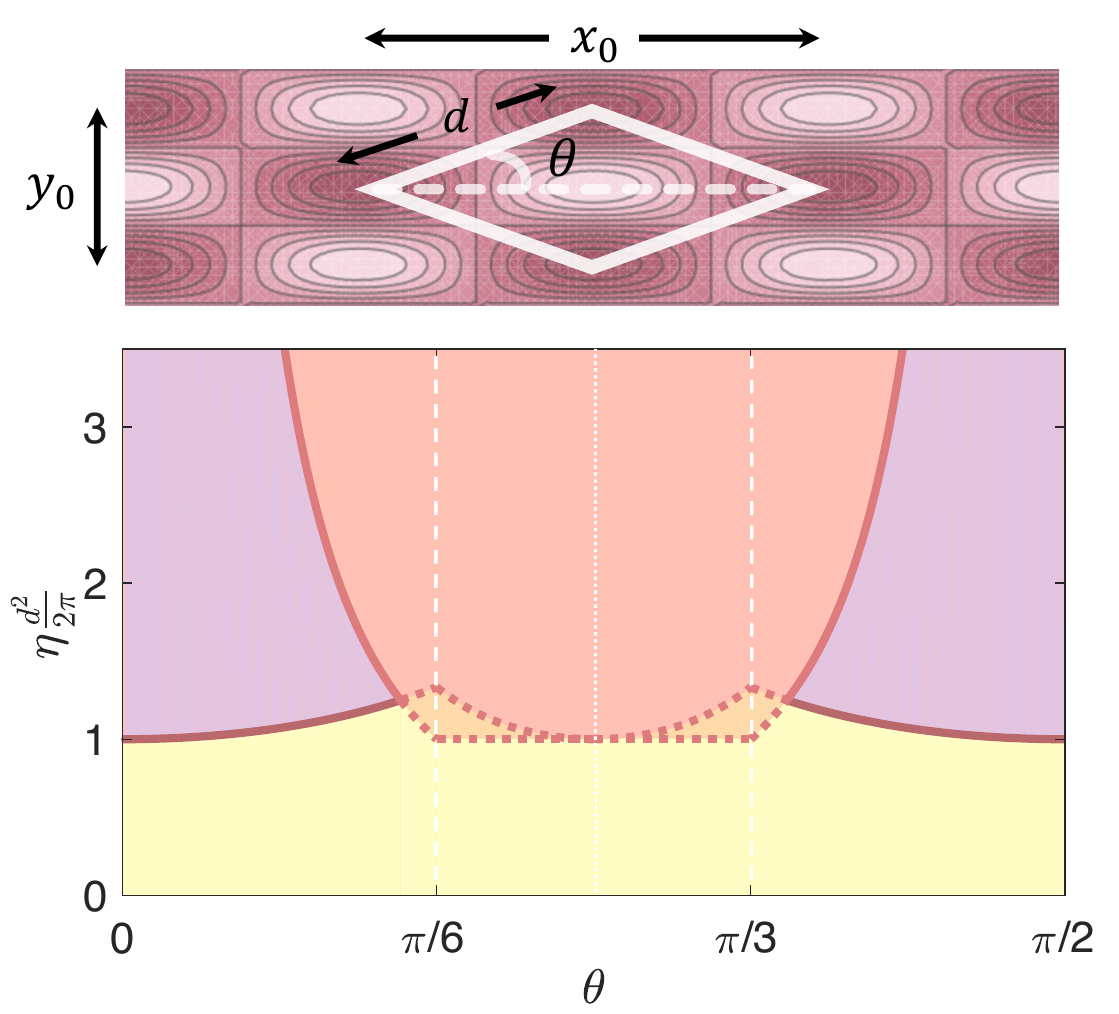}
\caption{\label{rh_pd} The phase boundaries for the 2D QBM model on centered rectangular periodic potentials are obtained from the flow diagrams in Fig.~\ref{rh_flow} and plotted as a function of the lattice parameter $\theta$ for fixed $d$, as denoted in the top schematic. The phase diagram contains regions of a single phase (delocalized (yellow), line-localized (purple), or point-localized (red)) and regions of two phases (orange), where the particle could be in the point-localized or the delocalized phase depending on the initial value of the potential $V_{xy}$. The dashed lines at $\theta = \pi/6$ and $\theta = \pi/3$ correspond to the triangular (hexagonal) lattice and the dotted line at $\theta = \pi/4$ corresponds to the square lattice.} 
\end{figure}

\begin{center}
\begin{table}[b]
\begin{tabular} {| c | c | c | c|}
\hline
	$\theta$ range & phases			&	$\frac{d^2}{2 \pi } \cdot \eta(\theta) $ \\
\hline
\hline
\multirow{4}{*}{$0<\theta< \tan^{-1}(\frac{1}{2})$} 	& \multirow{2}{*}{delocalized} & \multirow{3}{*}{------------ $\sec^2(\theta)$ ------------}\\
		&  \multirow{2}{*}{1D $x$ loc.} 	&  \multirow{3}{*}{------------ $ \frac{1}{4} \csc^2(\theta)$------------} \\
		&   \multirow{2}{*}{2D loc.} 			&  \\
		&&\\
\hline
\multirow{3}{*}{$\tan^{-1}(\frac{1}{2}) < \theta <\frac{\pi}{6} $} 		& \multirow{2}{*}{delocalized} & \multirow{3}{*}{----- $\frac{1}{4} \csc^2(\theta)$ to $ \sec^2(\theta)$ -----}  \\
		& \multirow{2}{*}{2D loc.} 			& \\
		&& \\
\hline
\multirow{3}{*}{$\frac{\pi}{6} < \theta <\frac{\pi}{3} $} 		& \multirow{2}{*}{delocalized} & \multirow{3}{*}{-----$1$ to $\frac{1}{4} \sec^2(\theta) \csc^2(\theta)$-----}  \\
		& \multirow{2}{*}{2D loc.} 			& \\
		&& \\
\hline
\multirow{3}{*}{$\frac{\pi}{3} < \theta <\cot^{-1}(\frac{1}{2}) $} 		& \multirow{2}{*}{delocalized} & \multirow{3}{*}{----- $\csc^2(\theta)$ to $\frac{1}{4} \sec^2(\theta)$ -----}  \\
		& \multirow{2}{*}{2D loc.} 			& \\
		&& \\
\hline
\multirow{4}{*}{$\cot^{-1}(\frac{1}{2}) < \theta<\frac{\pi}{2}$} 	& \multirow{2}{*}{delocalized} & \multirow{3}{*}{------------ $\csc^2(\theta)$ ------------}\\
		&  \multirow{2}{*}{1D $y$ loc. } 	&  \multirow{3}{*}{------------ $\frac{1}{4} \sec^2(\theta)$------------} \\
		&   \multirow{2}{*}{2D loc.} 			&  \\
		&&\\
\hline
\end{tabular}
{\caption {\label{rh_trans}} The possible phases exhibited by a 2D QBM model in a centered rectangular lattice potential, and the corresponding phase boundaries as a function of the lattice parameters, where $d$ and $\theta = \tan^{-1}(y_0/x_0)$ are as labeled in Fig.~\ref{rh_pd}. }
\end{table}
\end{center}

In QBM models on periodic potentials, self-duality exists between the real space and momentum space coordinates, $\bm{r}(\tau)$ and $\bm{k}(\tau)$. Therefore, the calculations performed in previous sections can be applied to the dual action, which describes the particle on its reciprocal lattice in momentum space. As shown previously in Refs.~\onlinecite{YiShort, YiLong}, the dual parameters are $V_{\bm{G}}$, the dimensionless amplitude for the potential at a reciprocal lattice vector ${\bm{G}}$, and $t_{\bm R}$, the dimensionless tunneling amplitude between two lattice sites connected by a displacement vector $\bm R$. The action in real space and its dual action in reciprocal space are then, 
\begin{equation}
\begin{aligned}
S &= \frac{1}{2} \int \frac{d\omega}{2\pi} \eta |\omega| |\bm{r}(\omega)|^2 - \int d\tau \sum_{\bm{G}} v_{\bm{G}} e^{i 2 \pi {\bm{G}} \cdot \bm{r}(\tau)} \\
S_{\text{dual}} &= \frac{1}{2} \int \frac{d\omega}{2\pi} \frac{1}{\eta} |\omega| |\bm{k}(\omega)|^2 - \int d\tau \sum_{\bm{R}} t_{\bm{R}} e^{i 2 \pi {\bm{R}} \cdot \bm{k}(\tau)}.
\end{aligned}
\end{equation}
This leads to the following flow equations,
\begin{equation} \label{scalingEq}
\begin{aligned}
\frac{\partial v_{\bm{G}}}{\partial l} = \Big( 1 - \frac{1}{\alpha_G} \Big)  v_{\bm{G}} \\
\frac{\partial t_{\bm{R}}}{\partial l} = \Big( 1 - {\bar \alpha_{\bm{R}}} \Big)  t_{\bm{R}}, \\
\end{aligned}
\end{equation}
where $\alpha_G = \eta \frac{1}{2 \pi |{\bm{G}}|^2}$, and $\bar \alpha_R = \eta 2 \pi |{\bm{R}}|^2 $. 

We can now take the analyses used in Sec.~\ref{rhSec} for centered rectangular lattices to study their corresponding reciprocal space lattices. In the system coupled to the dual reciprocal space lattice potential, the perturbative parameter becomes $t_{xy}$, the tunneling amplitude for a particle through the potential hills. For concreteness, we consider and explicitly draw out the flow diagrams for lattice potentials where the $x$-periodicity is longer than the $y$-periodicity ($x_0 > y_0$, where $x_0$ and $y_0$ are denoted as before in Fig.~\ref{rh_pd}). The real space lattice and its reciprocal space lattice are also drawn on the top of Fig.~\ref{rh_flow} for reference. Upon applying our analyses in Secs.~\ref{2Dnaive} and \ref{2Dredo} to the reciprocal space lattice and using Eq.~(\ref{scalingEq}), we arrive at the relevance conditions for the given potential and tunneling amplitude parameters and those generated under perturbative RG. These conditions are summarized in Tab.~\ref{dualTable} and used to construct the flow diagrams in Fig.~\ref{rh_flow}. Since $\eta$, $x_0$, and $y_0$ are not renormalized under RG, the phase transition lines and flow lines on each plane of the prism are straight, as in the case of the 1D periodic potential (Fig.~\ref{schm}). The zero temperature flow diagram for the 1D periodic potential shows that, in the regime where the dimensionless friction coefficient $\alpha <1$, any nonzero tunneling amplitude $t$ leads the flows to the fixed line at $t=\infty$, and in the regime where $\alpha > 1$, any nonzero potential barrier $V$ leads the flows to the fixed line at $V=\infty$. These arguments carry over precisely for each wall of the flow diagram prism in Fig.~\ref{rh_flow}. In other words, for a given value of $\eta$, we follow the flow lines in that section of the flow diagram prism to reach the corresponding fixed line, which determines the phase the system is exhibiting under those system parameters. 

Crucially, we show in Fig.~\ref{rh_flow} that the terminations of all flow lines are consistent under duality. We then obtain qualitatively distinct flow diagrams depending on the parameter $\theta$ of the lattice potential geometry. Flow diagrams corresponding to three ranges of $\theta$ where $x_0>y_0$ are shown in Fig.~\ref{rh_flow}. The corresponding behavior for $y_0 > x_0$ is then obtained by the exchange $x_0 \Leftrightarrow y_0$. We thus derive the phase boundaries of the zero temperature phase diagram in Fig.~\ref{rh_pd}, summarized in Tab.~\ref{rh_trans}, where $d$ and $\theta$ are still as denoted in the top of Fig.~\ref{rh_pd}. 

We see that for $\theta < \tan^{-1}(1/2)$, the system exhibits a line localized phase (purple in Fig.~\ref{rh_pd}), where the particle is localized in the direction of longer periodicity $x$, in between transitioning from the delocalized phase to the point localized phase. For ${\tan^{-1}(1/2) < \theta < \cot^{-1}(1/2)}$, the system can exhibit a phase (orange in Fig.~\ref{rh_pd}) where the particle can be in either the delocalized phase or the point localized phase. From the flow diagrams, there are two stable fixed points at $V_{xy} = 0$ and $V_{xy} = \infty$ and therefore at least one unstable fixed point between them. This is as in the case studied in previous literature for the triangular (hexagonal) lattice potential \cite{YiShort, YiLong}. As in previous analysis, we also assume the only stable fixed points are the two that we currently identify, and the zero-temperature phase in that region of the phase diagram depends on whether the initial value of $V_{xy}$ is above or below some unstable intermediate fixed point. 

\section{Conclusions}

In this paper, we have generalized the study of 2D QBM models to those coupled to four types of Bravais lattice potentials. The flow diagrams obtained from our analyses are consistent with the self-duality between the weak and strong coupling regimes. We have shown that anisotropy in the lattice symmetry is reflected in the type of phases the system is capable of exhibiting under the tuning of a phenomenological friction coefficient. Hence forth, calculations using the same methods may now be carried out to study phase transitions of QBM models in higher dimensional periodic potentials. Generalization of these calculations to lattice potentials of higher dimensions and different symmetries may be applicable, through mappings such as those mentioned in Sec.~\ref{intro}, to the studies of other systems in condensed matter and statistical physics. 


\begin{acknowledgments}
The author is very grateful to Mehran Kardar for his teaching, as well as invaluable discussions and comments on the paper. She also thanks Hengyun Zhou, Tomohiro Soejima, and Ruffin Evans for critical discussions and helpful advice. This material is based upon work supported by the National Science Foundation Graduate Research Fellowship under Grant No. DGE1745303.
\end{acknowledgments}


\appendix

\section{Perturbative RG analysis of the 2D centered rectangular potential} \label{apA}

First, we identify the two-point correlation functions of the bare action upon integrating over the fast modes $G_i(\tau)  \equiv \langle \phi_{i,>} (\tau) \phi_{i,>}(0) \rangle$,
\begin{equation}
\begin{aligned}
G_i(\tau) 	&= \frac{2}{\alpha_i} Y_0(\Lambda \tau /b) \quad \text{for } \Lambda\tau \gg 1, \\
G_i(0) &= \frac{2}{\alpha_i}\ln(b),
\end{aligned}
\end{equation}
where $i$ = $x$, $y$ and $\langle \phi_{x,> }(\tau) \phi_{y,>}(0) \rangle = 0$. 

Calculation is as follows,
\begin{equation}
\begin{aligned}
\langle S_1 \rangle 		&= V_{xy} \Lambda \int d\tau \langle \cos \big(\phi_{x, <} + \phi_{x,>} \big) \cos \big(\phi_{y,<} + \phi_{y,>} \big) \rangle \\ 
					&= V_{xy} \Lambda  \int d\tau \frac{1}{2} \Big[ \cos (\phi_{x,<} - \phi_{y,<}) \langle \cos (\phi_{x,>} - \phi_{y,>}) \rangle \\ 
					&\qquad   \qquad \qquad + \cos (\phi_{x,<} + \phi_{y,<}) \langle \cos (\phi_{x,>} + \phi_{y,>}) \rangle \Big] \\		
					&= V_{xy} \Lambda  \int d\tau e^{-\frac{1}{2}(G_x(0)+G_y(0))}\Big[ \cos (\phi_{x,<}) \cos( \phi_{y,<}) \Big] \\ 	
\end{aligned}
\end{equation}
Rescaling $\tau' = \tau b$ and linearizing $b \approx 1+\delta l$, we have that
\begin{equation}
\frac{\partial V_{xy}(l)}{\delta l} = \Bigg[1-\Big(\frac{1}{\alpha_x}+\frac{1}{\alpha_y}\Big) \Bigg] V_{xy}(l).
\end{equation}
Thus, $V_{xy}$ scales to 0 when $ \frac{\eta}{2 \pi } < \big( \frac{1}{x_0^2}+\frac{1}{y_0^2} \big)$. We check that when $x_0=y_0$, the imposed condition becomes identical to that of a square lattice with spacing $x_0/\sqrt{2}$. 
Physically, we expect the anisotropy of the potential to reflect in the system in some way, which motivates us to seek higher order effects. 
To minimize clutter in the following calculations, we will drop the subscript$_<$ label on the slow modes and transform notations for the fast modes as such $\cos(\phi_{x,>}(\tau')) \Rightarrow c_x'$. 
\small 
\begin{equation}
\begin{aligned} 
 \langle S_1^2 \rangle	&= (V_{xy} \Lambda)^2 \int d\tau d\tau' \langle (\cos(\phi_x) \cos(\phi_y) \cos(\phi_x') \cos(\phi_y') \rangle \\
		&= (V_{xy} \Lambda)^2 \int d\tau d\tau' \frac{1}{4}  \\
		\times \Big[ &\cos(\phi_{x}-\phi_{y})\cos(\phi_{x}'-\phi_{y}') \langle (c_x c_y + s_x s_y)(c'_x c'_y + s'_x s'_y) \rangle \\
		+ &\cos(\phi_{x}-\phi_{y})\cos(\phi_{x}'+\phi_{y}') \langle (c_x c_y + s_x s_y)(c'_x c'_y - s'_x s'_y) \rangle \\
		+ &\cos(\phi_{x}+\phi_{y})\cos(\phi_{x}'-\phi_{y}') \langle (c_x c_y - s_x s_y)(c'_x c'_y + s'_x s'_y) \rangle \\
		+ &\cos(\phi_{x}+\phi_{y})\cos(\phi_{x}'+\phi_{y}') \langle (c_x c_y - s_x s_y)(c'_x c'_y - s'_x s'_y) \rangle \Big].
\end{aligned}
\end{equation}
\normalsize
The second and third terms are identical upon exchange in $\tau \leftrightarrow \tau'$. The expectation values of the first and fourth terms are identical since cross terms such as $\langle c_x s_x \rangle$ has an argument that is odd in $\phi_x$ and thus vanish. The expectation values in the first and fourth terms are calculated to be, 
\small 
\begin{equation}
\begin{aligned}
&\langle (c_x c_y + s_x s_y)(c'_x c'_y + s'_x s'_y) \rangle = \langle (c_x c_y - s_x s_y)(c'_x c'_y - s'_x s'_y) \rangle \\
 		&\quad = e^{-(G_x(0)+G_y(0))} \cosh \Big[ G_x(\tau-\tau')+G_y(\tau-\tau')\Big]. \\
\end{aligned}
\end{equation}
\normalsize
The expectation values in the second and third term are calculated to be
\small
\begin{equation}
\begin{aligned}
&\langle (c_x c_y + s_x s_y)(c'_x c'_y - s'_x s'_y) \rangle \\
 		& \quad = e^{-(G_x(0)+G_y(0))}\cosh \Big[ G_x(\tau-\tau')-G_y(\tau-\tau')\Big]. \\
\end{aligned}
\end{equation}
\normalsize
The slow-mode function from the first and fourth terms are then collected to be
\small 
\begin{equation} \label{exp1}
\begin{aligned}
\cos&(\phi_{x}-\phi_{y})\cos(\phi_{x}'-\phi_{y}') + \cos(\phi_{x}+\phi_{y})\cos(\phi_{x}'+\phi_{y}') \\
=& \cos(\phi_x+\phi_x')\cos(\phi_y+\phi_y') + \cos(\phi_x-\phi_x')\cos(\phi_y-\phi_y'). 
\end{aligned}
\end{equation}
\normalsize
The slow-mode function from the second and third terms are collected to be
\small
\begin{equation} \label{exp2}
\begin{aligned}
2 &\cos(\phi_{x}-\phi_{y})\cos(\phi_{x}'+\phi_{y}')  \\
&= \cos(\phi_x+\phi_x')\cos(\phi_y-\phi_y') + \cos(\phi_x-\phi_x')\cos(\phi_y+\phi_y'),
\end{aligned}
\end{equation}
\normalsize
where in the collection of these terms, we have removed functions that are odd under the exchange of $\tau \leftrightarrow \tau'$. Approximating $\cos(\phi_i - \phi_i')$ by the gradient expansion and removing terms $\sim \big(\frac{\partial \phi_i}{\partial \tau}\big)^2$ by the same argument of power counting as before, we see that second order RG generates the following terms,
\begin{equation}
\begin{aligned}
V_{xy}^{(2)} \cos(2\phi_x) \cos(2 \phi_y) + V_x \cos(2\phi_x) + V_y \cos(2\phi_y), 
\end{aligned}
\end{equation}
where, using results (\ref{exp1}) and (\ref{exp2}), the coefficients are, at this order,
\begin{equation}
\begin{aligned}
V_{xy}^{(2)} &= \frac{(V_{xy}\Lambda)^2}{4}e^{-(G_x(0)+G_y(0))} \\
		&  \qquad \qquad \times \Big\{\cosh \Big[ G_x(\tau-\tau')+G_y(\tau-\tau')\Big]-1 \Big\} \\
V_x &= V_y = \frac{(V_{xy}\Lambda)^2}{4}e^{-(G_x(0)+G_y(0))} \\
		&  \qquad \qquad \times \Big\{\cosh \Big[ G_x(\tau-\tau')-G_y(\tau-\tau')\Big]-1 \Big\}. \\
\end{aligned}
\end{equation}

\section{RG analysis of single dimensional correlation function using the strong coupling expansion} \label{apB}

We examine the two-point correlation function $\langle \phi_x(0) \phi_x(0) \rangle$ under the strong coupling regime of $V_x$. From Eq.~(\ref{RGInst}) in the main text, we have that upon applying the rescaling and renormalization procedure,
\small
\begin{equation}
\begin{aligned}
	& \langle \phi_x(0) \phi_x(0) \rangle \\
=	& \frac{1}{Z}	\sum_N \Bigg[\frac{\big(\frac{\omega_0}{\sqrt{2}}\sqrt{\frac{2 s}{\pi}}e^{-s}\big)^N}{N!}\Bigg]^2 \sum_{\{\tau_i, e_i\}}  \int_{-\infty}^{\infty} d\tau_1\int_{\omega_0^{-1}}^{\infty} d\bar \tau_2 \ldots d\bar \tau_{2N}  \\
	&  \times \prod_{j\neq k}^{2N} \Big(\frac{\omega_0}{\sqrt{2}} (\tau_j-\tau_k)\Big)^{\frac{\alpha_x}{4} e_j e_k } \sum_{m, n=1}^{2N} e_m e_n f(-b^2 \tau_m) f(-b^2 \tau_n). 
\end{aligned}
\end{equation}
\normalsize
We look at each term in the $N$-summation individually as follows. 
First, we define and study the following expression $C_{N}(\alpha_x, b)$, the motivation for which will become clear later,
\begin{widetext}
\begin{equation}
C_{N}(\alpha_x, b) = \frac{\int_{-\infty}^{\infty} d\tau_1\int_{\omega_0^{-1}}^{\infty} d\bar \tau_2 \ldots d\bar \tau_{2N}  \prod_{j\neq k}^{2N} \Big(\frac{\omega_0}{\sqrt{2}} (\tau_j-\tau_k)\Big)^{\frac{\alpha_x}{4} e_j e_k } \sum_{m, n=1}^{2N} e_m e_n f(-b^2 \tau_m) f(-b^2 \tau_n) }{\int_{-\infty}^{\infty} d\tau_1\int_{\omega_0^{-1}}^{\infty} d\bar \tau_2 \ldots d\bar \tau_{2N}  \prod_{j\neq k}^{2N} \Big(\frac{\omega_0}{\sqrt{2}} (\tau_j-\tau_k)\Big)^{\frac{\alpha_x}{4} e_j e_k }}
\end{equation}
\end{widetext}
We first write out this expression for the $N=1$ term. The following expression then corresponds to both sequences $\{(+, -), (-, +)\}$, exhausting all possible configurations for a 2-particle neutral instanton gas. 
\begin{widetext}
\begin{equation}
\begin{aligned}
C_{1}(\alpha_x, b) = & \frac{\int_{-\infty}^{\infty} d\tau_1\int_{\omega_0^{-1}}^{\infty} d\bar \tau_2 \Big(\frac{\omega_0}{\sqrt{2}} (\tau_2-\tau_1)\Big)^{-\frac{\alpha_x}{2} } (f(-b^2 \tau_1)-f(-b^2 (\tau_1+\bar \tau_2)))^2 }{\int_{-\infty}^{\infty} d\tau_1\int_{\omega_0^{-1}}^{\infty} d\bar \tau_2  \Big(\frac{\omega_0}{\sqrt{2}} (\tau_2-\tau_1)\Big)^{-\frac{\alpha_x}{2} }} \\
= & \frac{\int_{\omega_0^{-1}}^{\infty} d\bar \tau_2 \Big(\frac{\omega_0}{\sqrt{2}} \bar \tau_2\Big)^{-\frac{\alpha_x}{2} } (f(0)-f(-b^2 \bar \tau_2))^2 }{\int_{\omega_0^{-1}}^{\infty} d\bar \tau_2  \Big(\frac{\omega_0}{\sqrt{2}} \bar \tau_2\Big)^{-\frac{\alpha_x}{2} }} \\
= & \frac{\int_{\omega_0^{-1}}^{\infty} d\bar \tau_2 \Big(\frac{\omega_0}{\sqrt{2}} \bar \tau_2\Big)^{-\frac{\alpha_x}{2} } (\pi-4\tan^{-1}(e^{-b^2 \omega_0 \bar \tau_2}))^2 }{\int_{\omega_0^{-1}}^{\infty} d\bar \tau_2  \Big(\frac{\omega_0}{\sqrt{2}} \bar \tau_2\Big)^{-\frac{\alpha_x}{2} }}  \\ 
= & \frac{\int_{1}^{\infty} d\rho \quad \rho^{-\frac{\alpha_x}{2} } (\pi-4\tan^{-1}(e^{-b^2 \rho}))^2 }{\int_{1}^{\infty} d\rho \quad \rho^{-\frac{\alpha_x}{2} }}.  \label{N1Int}
\end{aligned}
\end{equation}
\end{widetext}
In the second line, we have cancelled out the integration over $\tau_1$, the degree of freedom that indicates where an instanton gas is located in absolute time space. By time invariance, we simply take $\tau_1=0$ for calculations. In the last line, we have rescaled $\rho = \omega_0 \tau$ to be dimensionless so that the integral no longer depends on $\omega_0$. 
We also explicitly write out the $N=2$ terms for concreteness,
\small
\begin{widetext}
\begin{equation} \label{N2AConfig}
\begin{aligned}
C_{2}^{(1)}(\alpha_x, b) = &\frac
{\int_{1}^{\infty} d\rho_2 d\rho_3 d\rho_4 \quad \Big[ \frac{\rho_2 \rho_4 (\rho_2+\rho_3) (\rho_3+\rho_4) }{\rho_3 (\rho_2+\rho_3+\rho_4)} \Big]^{-\frac{\alpha_x}{2}}
 (\pi-4\tan^{-1}(e^{-b^2 \rho_2})-4\tan^{-1}(e^{-b^2 (\rho_2+\rho_3)})+4\tan^{-1}(e^{-b^2 (\rho_2+\rho_3+\rho_4)}))^2 }
{\int_{1}^{\infty} d\rho_2  d\rho_3 d\rho_4 \quad \Big[ \frac{\rho_2 \rho_4 (\rho_2+\rho_3) (\rho_3+\rho_4) }{\rho_3 (\rho_2+\rho_3+\rho_4)} \Big]^{-\frac{\alpha_x}{2}}}, \\
\end{aligned}
\end{equation}

\begin{equation} \label{N2BConfig}
\begin{aligned}
C_{2}^{(2)}(\alpha_x, b) = &\frac
{\int_{1}^{\infty} d\rho_2 d\rho_3 d\rho_4 \quad \Big[ \frac{\rho_2 \rho_3  \rho_4 (\rho_2+\rho_3+\rho_4)}{ (\rho_3+\rho_4) (\rho_2+\rho_3)} \Big]^{-\frac{\alpha_x}{2}}
 (\pi-4\tan^{-1}(e^{-b^2 \rho_2})+4\tan^{-1}(e^{-b^2 (\rho_2+\rho_3)})-4\tan^{-1}(e^{-b^2 (\rho_2+\rho_3+\rho_4)}))^2 }
{\int_{1}^{\infty} d\rho_2 d\rho_3 d\rho_4 \quad \Big[ \frac{\rho_2 \rho_3  \rho_4 (\rho_2+\rho_3+\rho_4)}{ (\rho_3+\rho_4) (\rho_2+\rho_3)} \Big]^{-\frac{\alpha_x}{2}}}, \\
\end{aligned}
\end{equation}

\begin{equation} \label{N2CConfig}
\begin{aligned}
C_{2}^{(3)}(\alpha_x, b) = &\frac
{\int_{1}^{\infty} d\rho_2 d\rho_3 d\rho_4 \quad \Big[ \frac{\rho_3 (\rho_2+\rho_3) (\rho_3+\rho_4) (\rho_2+\rho_3+\rho_4)  }{\rho_2 \rho_4 } \Big]^{-\frac{\alpha_x}{2}}
 (\pi+4\tan^{-1}(e^{-b^2 \rho_2})-4\tan^{-1}(e^{-b^2 (\rho_2+\rho_3)})-4\tan^{-1}(e^{-b^2 (\rho_2+\rho_3+\rho_4)}))^2 }
{\int_{1}^{\infty} d\rho_2 d\rho_3 d\rho_4 \quad \Big[ \frac{\rho_3 (\rho_2+\rho_3) (\rho_3+\rho_4) (\rho_2+\rho_3+\rho_4)  }{\rho_2 \rho_4 } \Big]^{-\frac{\alpha_x}{2}}}, \\
\end{aligned}
\end{equation}
\end{widetext}
\normalsize
where Eq.~(\ref{N2AConfig}) corresponds to the sequences $\{(+, -, +, -), (-, +, -, +)\}$, Eq.~(\ref{N2BConfig}) corresponds to the sequences $\{(+, -, -, +), (-, +, +, -)\}$, and Eq.~(\ref{N2CConfig}) corresponds to the sequences $\{(+, +, -, -), (-, -, +, +)\}$, exhausting all possible distinct configurations for a 4-particle neutral instanton gas. 

Upon looking at the explicit expressions for each $C_N(\alpha_x, b)$, it becomes clear that as we take the limit of $b \rightarrow \infty$, each term after the initial term in the integrand of the numerator vanishes. The remaining integrals in the numerator and denominator then cancel out, and we are left with a finite constant of $\pi^2$. Since this is true for $C_N(\alpha_x, b)$ for all values of $N$, the constant $\pi^2$ can be taken out of the summation in $N$, and the expression for $G_x(0)$ must approach the same limit:
\begin{equation}
\lim_{b\rightarrow \infty} G_x(0) = \lim_{b\rightarrow \infty} C(\alpha_x, b) = \pi^2.
\end{equation}

\bibliography{8334_QF_Edit1}

\end{document}